\documentclass[a4paper,11pt]{article}
\usepackage{jheppub,slashed,color,graphicx,subfigure,dsfont,slashed,multirow,lscape,hyperref}
\usepackage[utf8]{inputenc}

\renewcommand{\arraystretch}{1.2}
\newcommand{\GeV}{{\rm ~GeV}}
\newcommand{\TeV}{{\rm ~TeV}}
\newcommand{\fb}{{\rm ~fb}}
\newcommand{\invfb}{{\rm ~fb^{-1}}}
\newcommand{\BR}[1]{{{\rm BR}\left(#1\right)}}
\newcommand{\vareps}{\varepsilon}
\newcommand{\WR}{W_R}
\newcommand{\ZR}{Z_R}
\newcommand{\Wssmpm}{W_{\rm SSM}^\pm}
\newcommand{\Wssmp}{W_{\rm SSM}^+}
\newcommand{\Wssm}{W_{\rm SSM}}
\newcommand{\Zssm}{Z_{\rm SSM}}
\newcommand{\Gam}[1]{\Gamma\left(#1\right)}
\newcommand{\MWprime}{M_{W_{\rm SSM}}}
\newcommand{\pTVeto}{p_T^{\rm Veto}}
\newcommand{\MET}{\slashed{E}_T}
\newcommand{\tWSq}{\tan^2\theta_W}
\newcommand{\mg}{MG5aMC}
\newcommand{\mgLong}{{\sc MadGraph5}\_a{\sc MC@NLO}}
\newcommand{\be}{\begin{equation}}
\newcommand{\ee}{\end{equation}}
\def\bsp#1\esp{\begin{split}#1\end{split}}

\newcommand{\comment}{\textcolor{black}}
\newcommand{\confirm}{\textcolor{black}}

\title{A Comprehensive Framework for Studying $W'$ and $Z'$ Bosons at Hadron
  Colliders with Automated Jet Veto Resummation}

\author[a,b,c]{Benjamin Fuks}
\author[d]{and Richard Ruiz}

\emailAdd{fuks@lpthe.jussieu.fr}
\emailAdd{richard.ruiz@durham.ac.uk}

\affiliation[a]{Sorbonne Universit\'es, UPMC Univ.~Paris 06, UMR 7589, LPTHE,
  F-75005, Paris, France}
\affiliation[b]{CNRS, UMR 7589, LPTHE, F-75005, Paris, France}
\affiliation[c]{Institut Universitaire de France, 103 boulevard Saint-Michel,
  75005 Paris, France}
\affiliation[d]{Institute for Particle Physics Phenomenology {\rm(IPPP)},
Department of Physics, Durham University, Durham, DH1 3LE, UK}

\abstract{
The production of high-mass, color-singlet particles in hadron collisions is universally accompanied by initial state QCD radiation that is predominantly 
soft with respect to the hard process scale $Q$ and/or collinear with respect to the beam axis.
At TeV-scale colliders, this is in contrast to top quark and multijet processes, which  are hard and central.
Consequently, vetoing events with jets possessing transverse momenta above $p_T^{\rm Veto}$ in searches for new color-singlet 
states can efficiently reduce non-singlet backgrounds, thereby increasing experimental sensitivity.
To quantify this generic observation, we investigate the production and leptonic decay of a Sequential Standard Model $W'$ 
boson at the 13 TeV Large Hadron Collider.
We systematically consider signal and background processes at next-to-leading-order (NLO) in QCD with parton shower (PS) matching.
For color-singlet signal and background channels, we resum Sudakov logarithms of the form $\alpha_s^j(p_T^{\rm Veto})\log^k(Q/p_T^{\rm Veto})$
up to next-to-next-to-leading logarithmic accuracy (NNLL) with NLO matching.
We obtain our results using the {\sc MadGraph5\_aMC@NLO} and {\sc MadGraph5\_aMC@NLO-SCET} frameworks, respectively.
Associated Universal {\sc FeynRules} Output model files capable of handling NLO+PS- and NLO+NNLL-accurate computations are publicly available.
We find that within their given uncertainties, both the NLO+PS and NLO+NNLL(veto) calculations give accurate and consistent predictions.
Consequently, jet vetoes applied to color-singlet processes can be reliably modeled  at the NLO+PS level.
With respect to a $b$-jet veto of $p_{T}^{\rm Veto} = 30$ GeV, 
flavor-agnostic jet vetoes of $p_{T}^{\rm Veto} = 30-40$ GeV can further reduce single top and
$t\overline{t}$ rates by a factor of 2-50  at a mild cost of the signal rate.
Jet vetoes can increase the signal-to-noise ratios by roughly 10\% for light $W'$ boson masses of $30-50$~GeV and
25\%-250\% for masses of 300-800~GeV.
}

\begin{document}
\preprint{IPPP/17/1}

\maketitle
\flushbottom

\section{Introduction}\label{sec:intro}
The existence new massive, colorless vector bosons is a key prediction of many theories
that address the empirical and theoretical shortcomings of the Standard Model
(SM) of particle physics. This, for instance, includes dark photons and $Z'_D$
bosons in dark matter models, $W_{R}^\pm$ and $Z_R$ gauge bosons in left-right
symmetric models, $Z'_{B-L}$ bosons in neutrino mass models, or $W_{KK}^\pm$,
$Z_{KK}$ and $\gamma_{KK}$ Kaluza-Klein excitations in extra-dimension models.
These bosons are generically referred to as $W^{'\pm}$ and $Z'$
bosons~\cite{Altarelli:1989ff}.
Searches for these particles are and will continue to be an
integral part of the Large Hadron Collider (LHC) physics program. 
Subsequently, an ability to categorically increase the experimental sensitivity of such searches is desirable.

\begin{figure}
  \centering
  \includegraphics[width=.48\textwidth]{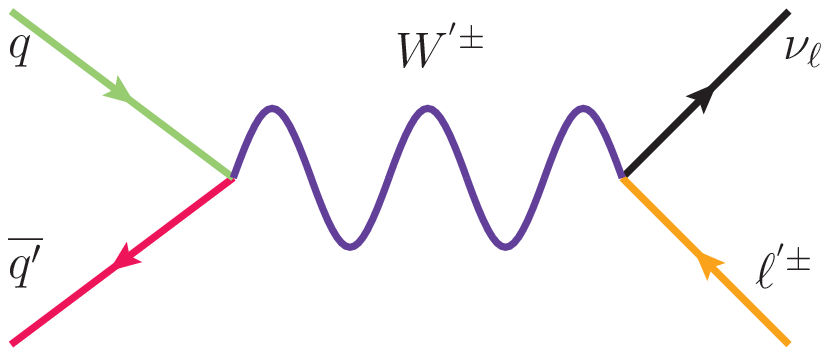}
  \includegraphics[width=.48\textwidth]{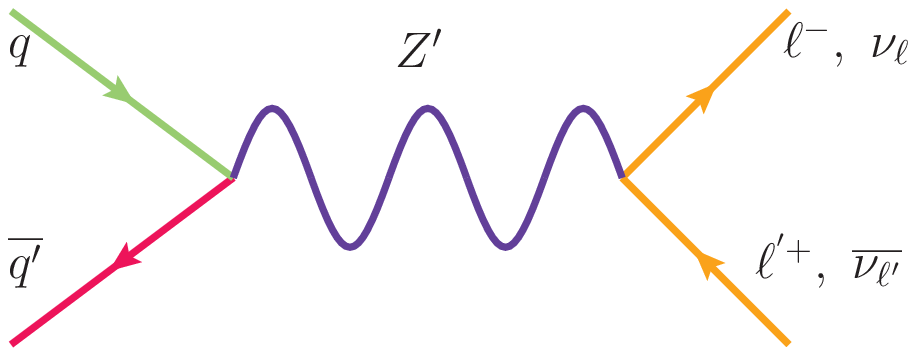}
  \caption{Born-level Feynman diagrams for $W'$ boson (left) and $Z'$-boson
    (right) production and decay into leptons in hadronic collisions.
    All figures are drawn with {\sc JaxoDraw}~\cite{Binosi:2003yf}.}
  \label{fig:production}
\end{figure}

A powerful and robust LHC test of these models consists of reinterpreting searches for leptonic decays
of Sequential Standard Model (SSM) $\Wssmpm$ and $\Zssm$ 
bosons~\cite{ATLAS:2016cyf,CMS:2016abv,ATLAS:2016ecs,CMS:2015kjy,CMS:2016ppa,Khachatryan:2016jww}, 
which proceed through the Drell-Yan (DY) processes
\be
 q \overline{q'} \to \Wssmpm \to \ell^\pm \nu_\ell
 \ \ \text{and}\ \
 q \overline{q} \to \Zssm \to \ell^+ \ell^-
   \ \ {\rm with~} q,q'\in\{u,d,c,s,b\}~{\rm and~}\ell \in \{e,\mu,\tau\}\ ,
\label{eq:ssmProcess} \ee
and whose leading order (LO) diagrams are shown in Figure~\ref{fig:production}.
SSM bosons couple to SM fermions in the same manner as the SM $W^\pm$
and $Z$ bosons up to overall coupling normalizations and may thus possess
couplings that radically differ from any of the aforementioned models. It is
nonetheless straightforward to reinterpret the SSM collider limits on masses and
couplings within another theoretical framework.

The collider signatures relevant for the benchmark processes of
Eq.~\eqref{eq:ssmProcess} consist of final states made of one or two
charged leptons ($\ell^\pm$) with a large transverse momentum ($p_T$),
and additionally accompanied by a large amount of missing transverse
energy ($\MET$) in the $\Wssm$ case. The dominant (and irreducible) backgrounds
are thus the charged current and neutral current DY
continua. However, as in most hadron collider searches, SSM searches are
\textit{inclusive} with respect to jet and soft QCD activity, so that
high-$p_T$ multijet and top quark processes, with cross sections many orders of
magnitude larger than the SSM boson production rates, also contribute to the SM
background.

Intriguingly, $s$-channel $\Wssm$ and $\Zssm$ production are color-singlet
$q\overline{q}$ annihilation processes, meaning that QCD radiation off the initial-state quarks is
favorably soft with respect to the hard process scale $Q\sim M_{W'/Z'}$
and/or collinear with respect to the beam axis.
This implies that the corresponding
jet activity is inherently softer and more collinear than for the leading QCD
backgrounds. In particular, top quark decay products characteristically feature
large momenta scales of $p_T\sim50-60\GeV$, which suggests that inclusive,
{\it i.e.}, flavor-independent, jet vetoes,
even loose ones, can enhance signal-to-noise ratios in SSM boson searches. 

Historically, such arguments were made first for the vector boson fusion (VBF) process~\cite{Barger:1991ar}.
More recently, they have also been discussed in 
DY, non-VBF Higgs and multi-boson production 
channels~\cite{Jaiswal:2015vda,Dawson:2012gs,Banfi:2012yh,Becher:2012qa,Becher:2013xia,
Stewart:2013faa,Tackmann:2012bt,Berger:2010xi,Stewart:2010pd,Stewart:2009yx,
Dawson:2016ysj,Banfi:2012jm,Banfi:2015pju,Jaiswal:2014yba,Monni:2014zra,Meade:2014fca,Becher:2014aya},
as well as in several searches for physics beyond the Standard Model (BSM)~\cite{Ebert:2016idf,Aad:2012fw,Aad:2014vka,Tackmann:2016jyb}.
We argue, however, that due to the universality of QCD radiation in the soft and collinear limits,
jet vetoes are in fact generically applicable to \textit{any} color-singlet process 
that results in colorless final-state particles and 
in which QCD processes are a non-negligible fraction of the background.

As learned from measurements of the (inclusive) $W^+W^-+0j$ cross section at the 
LHC~\cite{Khachatryan:2015sga,Chatrchyan:2013oev,Chatrchyan:2013yaa,ATLAS:2012mec,
Jaiswal:2014yba,Monni:2014zra,Meade:2014fca,Becher:2014aya},
reliable predictions for color-singlet processes initiated by $q\overline{q}$ annihilation and gluon fusion 
on which a jet veto of scale $\pTVeto$ is applied require 
the resummation of Sudakov logarithms of the form $\alpha_s(\pTVeto)\log(Q/\pTVeto)$. 
These logarithms originate from $t$-channel propagators corresponding to
initial-state gluon radiation and spoil the convergence of the perturbative
series when the hard process mass scale $Q$ is much greater than the veto scale $\pTVeto$.
Indeed, the lowest order at which the $p_T$ spectrum of any color-singlet system 
is even qualitatively accurate everywhere is at next-to-leading order (NLO) in
QCD for the inclusive process matched to the leading logarithmic (LL) resummation of the recoiling radiation momenta~\cite{Collins:1985ue}.
Formally, this is the same accuracy as inclusive NLO calculations matched with presently available parton shower (PS).
Recently, jet veto resummation at
next-to-next-to-leading (NNLL) logarithmic accuracy has been
automated~\cite{Becher:2014aya}. In the latter case, resummed computations for
generic color-singlet processes are matched with the NLO
fixed order results within the \mgLong~(\mg)~platform~\cite{Alwall:2014hca} and the
Soft-Collinear Effective Theory~\cite{Bauer:2000yr,Bauer:2001yt,Beneke:2002ph} (SCET) formalism.

In this study, we investigate the impact of (resummed) jet vetoes on current and
future searches for generic $W'$ and $Z'$ bosons both at the LHC and at hypothetical
very large hadron colliders~\cite{Golling:2016gvc,Arkani-Hamed:2015vfh}, such as the proposed
Future Circular Collider (FCC) at CERN~\cite{FCC} or the Circular $pp$ Collider (CppC) at IHEP~\cite{CppC}. 
We focus, as a benchmark scenario, on the case of a $\Wssm$ boson, 
and perform our study on the basis of the automated resummation technology implemented within the
\mg+SCET framework.
{We compare our veto resummed results to those obtained at NLO+PS accuracy.}
Our work includes the construction of a new general
{\sc FeynRules}~\cite{Alloul:2013bka} model for extra gauge boson studies
that can be interfaced with {\sc NloCT}~\cite{Degrande:2014vpa} and {\sc FeynArts}~\cite{Hahn:2000kx}.
The associated Universal {\sc FeynRules} Output (UFO)~\cite{Degrande:2011ua} model
is publicly available from the {\sc FeynRules} model database~\cite{WprimeZprimeAtNLO} and
can be used to simulate hadronic and leptonic collisions up to NLO+PS accuracy.

The remainder of this study continues in the following order: 
In Section~\ref{sec:theory}, 
we present the theoretical framework for extending the SM field content extra gauge boson,
summarize current LHC constraints on new vector bosons, and provide details on our computational setup.
In Section~\ref{sec:vetoResum}, we briefly review jet veto resummation within
the SCET formalism,
and discuss {rate uncertainties associated with mass,} $\pTVeto$, and jet radius scale choices in $W'$ boson production.
Signal and background modeling with jet vetoes is described in
Section~\ref{sec:phenoModeling}, and we focus particularly in
Section~\ref{sec:metJetModeling} on the modeling of the missing energy and jet
properties in $W'\rightarrow e/\mu+\MET$ searches.
We then dedicate 
Section~\ref{sec:observability} to estimating the improved discovery for $\Wssm$ boson searches
gained by applying jet vetoes and finally
summarize and conclude in Section~\ref{sec:summary}.

\section{An effective framework for $W'$ and $Z'$ studies at colliders}\label{sec:theory}
\subsection{A Simplified Model for Extra Gauge Boson Searches}

We take a simplified approach to modeling physics beyond the Standard Model.
We do this by minimally extending the SM to construct a general effective framework
that can be used for studying various models featuring extra colorless gauge bosons
that couple to SM fermions.
Specifically, we supplement the SM field content by two
massive, colorless vector fields $W^{\prime\pm}$ and $Z'$ that are respectively
electrically charged and neutral.
To ensure model independence, 
the exact form of the $W'$ and $Z'$ chiral couplings to SM fermions is not specified,
and any interaction of the new vector bosons with other gauge or scalar bosons is omitted.
Following~Ref.~\cite{Gopalakrishna:2010xm,Han:2012vk}, the Lagrangian parameterizing the new vector bosons' couplings
to up-type and down-type quark fields $u_i$ and $d_j$ is given by 
\be\bsp
 \mathcal{L}^q_{\rm NP} =&\
    - \frac{g}{\sqrt{2}} \sum_{i,j} \Big[
       \overline{u}_i V_{ij}^{\rm CKM}~W_\mu^{\prime+}~\gamma^\mu
       \Big(\kappa_L^{q} P_L + \kappa_R^{q}P_R\Big)~ d_j + \text{H.c.}\Big]\\ &\qquad
    - \frac{g}{\cos\theta_W} \sum_{q=u,d}\sum_i \Big[
      \overline{q}_i ~Z_\mu^\prime \gamma^\mu
      \Big(\zeta_{L}^{q}P_L + \zeta_{R}^{q}P_R\Big)q_i\Big] \ ,
\esp\label{eq:LagWp}\ee
where $i$ and $j$ denote flavor indices, $P_{L/R}=\frac12 (1\mp\gamma_5)$ 
are the usual left/right-handed
chirality projectors, $V^{\rm CKM}$ is the Cabbibo-Kobayashi-Maskawa (CKM) matrix, 
and $g$ and $\theta_W$ are the weak coupling constant and mixing angle respectively. 
We choose coupling normalizations facilitating the mapping to the reference SSM
Lagrangian ${\cal L}_{\rm SSM}$~\cite{Altarelli:1989ff}.
The real-valued quantities $\kappa_{L,R}^q$ and $\zeta_{L,R}^q$
serve as overall normalization of the new interactions relative to the strength of the weak coupling constant. 
We do not assume additional sources of flavor violation beyond the SM CKM matrix.

Similarly, the interactions involving charged lepton $\ell$ and massless neutrino $\nu_\ell$ fields
are parametrized by~\cite{Gopalakrishna:2010xm,Han:2012vk}
\be\bsp
  \mathcal{L}^\ell_{\rm NP} = &\
   - \frac{g}{\sqrt{2}}\sum_i\Big[
       \overline{\nu}_{\ell_i} W_\mu^{\prime+}~\gamma^\mu
        \kappa_L^{\ell} P_L\ell_i^- + \text{H.c.}\Big]\\ &\qquad
    - \frac{g}{\cos\theta_W} \sum_{f=\ell,\nu_\ell}\sum_i \Big[
      \overline{f}_i ~Z_\mu^\prime \gamma^\mu
      \Big(\zeta_{L}^{f}P_L + \zeta_{R}^{f}P_R\Big)f_i\Big] \ .
\esp\label{eq:LagZp}\ee
The quantities $\kappa_L^\ell$ are real-valued and serve as normalizations for 
leptonic coupling strengths. 
As no right-handed neutrinos are present in the SM, the corresponding
right-handed leptonic new physics couplings are omitted ($\zeta_R^\nu =
\kappa_R^\ell = 0$).
We assume that leptonic interactions are flavor diagonal.

\begin{table}
\renewcommand{\arraystretch}{2}
\setlength\tabcolsep{6pt}
\centering
\begin{tabular}{ c || c | c | c | c || c | c | c  }
Charge & $u_L$ & $d_L$ & $\nu_L$ & $e_L$ & $u_R$ & $d_R$ & $e_R$\\
\hline \hline
$T_{L}^{3,f}$ & $+\frac12$ & $-\frac12$ & $+\frac12$ & $-\frac12$ & 0 & 0 & 0 \\
\hline
$Q^f$ & $+\frac23$ & $-\frac13$ & $0$ & $-1$ & $+\frac23$ & $-\frac13$ & $-1$
\end{tabular}
\caption{Weak isospin and electric charge assignments for the left-handed and
  right-handed chiral fermions $f_L$ and $f_R$ entering the $\Zssm$ vector and axial-vector couplings of
  Eq.~\eqref{eq:zVACoup}.} \label{tb:qNumbers}
\end{table}

From our general Lagrangians of Eq.~\eqref{eq:LagWp} and Eq.~\eqref{eq:LagZp},
the SSM limit is obtained by imposing the coupling strengths to be equal to the
SM weak couplings up to an overall normalization factor,
\be
  \zeta_{R,L}^{f} = \zeta_{\Zssm}^f \Big( g_V^f \pm g_A^f\Big)
  \quad\text{with}\quad
  g_V^f = \frac{1}{2}T_L^{3,f} - Q^f\sin^2\theta_W
  \quad\text{and}\quad
  g_A^f=-\frac{1}{2}T_L^{3,f},
\label{eq:zVACoup} \ee
where the quantum number assignments are listed in Table~\ref{tb:qNumbers}. In
the canonical SSM, the overall normalizations are further trivially fixed as
\be
 \kappa^{q,\ell}_L=1, \qquad
 \kappa_R^q=0,
 \qquad\text{and}\qquad
 \zeta_{\Zssm}^{f}=1 \ .
\label{eq:ssmParamNorm} \ee

\begin{table}
\renewcommand{\arraystretch}{2}
\setlength\tabcolsep{6pt}
\centering
\begin{tabular}{ c | c || c | c | c | c || c | c | c | c }
Gauge group & Charge & $u_L$ & $d_L$ & $\nu_L$ & $e_L$ & $u_R$ & $d_R$ & $N_R$
   & $e_R$\\
  \hline\hline
  $SU(2)_L$ & $T_{L}^{3,f}$ & $+\frac12$ & $-\frac12$ & $+\frac12$ & $-\frac12$
     & 0 & 0 & 0 & 0\\
  \hline
  $SU(2)_R$ & $T_{R}^{3,f}$ & 0 & 0 & 0 & 0 & $+\frac12$ & $-\frac12$ &
     $+\frac12$ & $-\frac12$\\
  \hline
  $U(1)_{\rm EM}$ & $Q^f$ & $+\frac23$ & $-\frac13$ & $0$ & $-1$ &
     $+\frac23$ & $-\frac13$ & $0$ & $-1$
\end{tabular}
\caption{Weak isospin and electric charge assignments for the left-handed and
  right-handed chiral fermions $f_L$ and $f_R$ entering the $\ZR$ couplings of
  Eq.~\eqref{eq:ZRcoup}. Right-handed neutrinos $N_R$ are included for
  completion.} \label{tb:qNumbers2}
\end{table}

This parameterization can be used to describe any model featuring extra
colorless, massive vector bosons, 
provided there is no new source of flavor violation with respect to the SM. 
For instance, right-handed $\WR$ and $\ZR$ boson interactions can be
obtained by enforcing
\be\bsp
  \zeta_L^f =&\ \frac{\kappa_{R}^f\cos\theta_W}
    {\sqrt{1 - \frac{\tWSq}{\big(\kappa_{R}^f\big)^2}}}
    \frac{\tWSq}{\big(\kappa_{R}^{f}\big)^2}\ \Big[T_L^{3,f} - Q^f\Big]\\
  \zeta_R^f =&\ \frac{\kappa_{R}^f\cos\theta_W}
    {\sqrt{1 - \frac{\tWSq}{\big(\kappa_{R}^f\big)^2}}}
    \Big[T_{R}^{3,f} - \frac{1}{\kappa_{R}^{f~2}}\tWSq Q^f\Big]\ ,
\esp\label{eq:ZRcoup}\ee
with $\kappa_{R}^{q,\ell}$ being the free parameters entering the interactions
of Eq.~\eqref{eq:LagWp} (in which $\kappa_L^{q,\ell} = 0$) and where the
electric and isospin charges are shown in Table~\ref{tb:qNumbers2}. Right-handed
neutrino couplings could be easily added in our effective framework, following
the minimal parameterization of Ref.~\cite{Mattelaer:2016ynf}.

\begin{figure}
  \centering
  \subfigure[]{
    \includegraphics[scale=1,width=.47\textwidth]{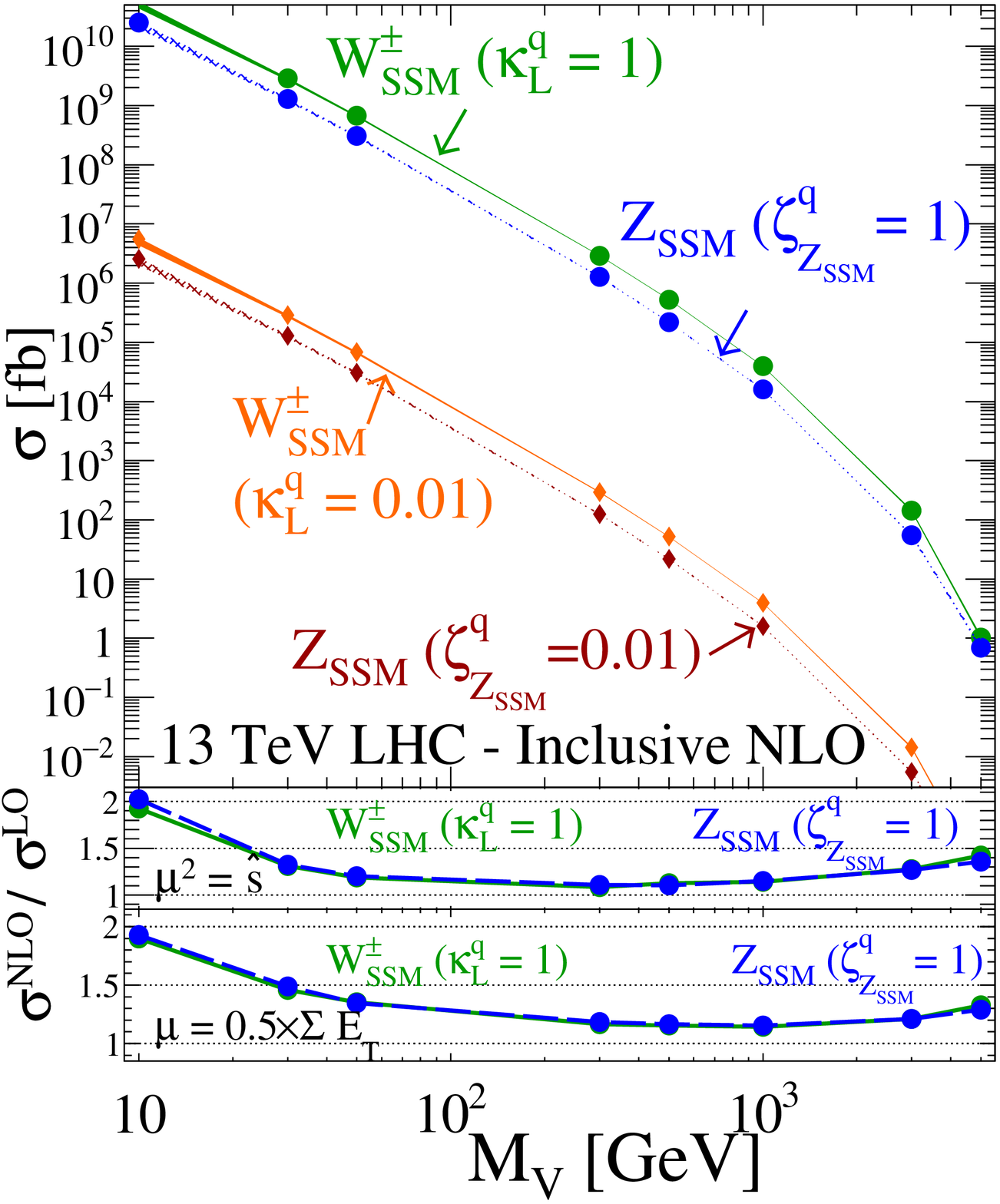}
    \label{fig:xsec13TeV}
  }
  \hspace{.25cm}\subfigure[]{
    \includegraphics[scale=1,width=.47\textwidth]{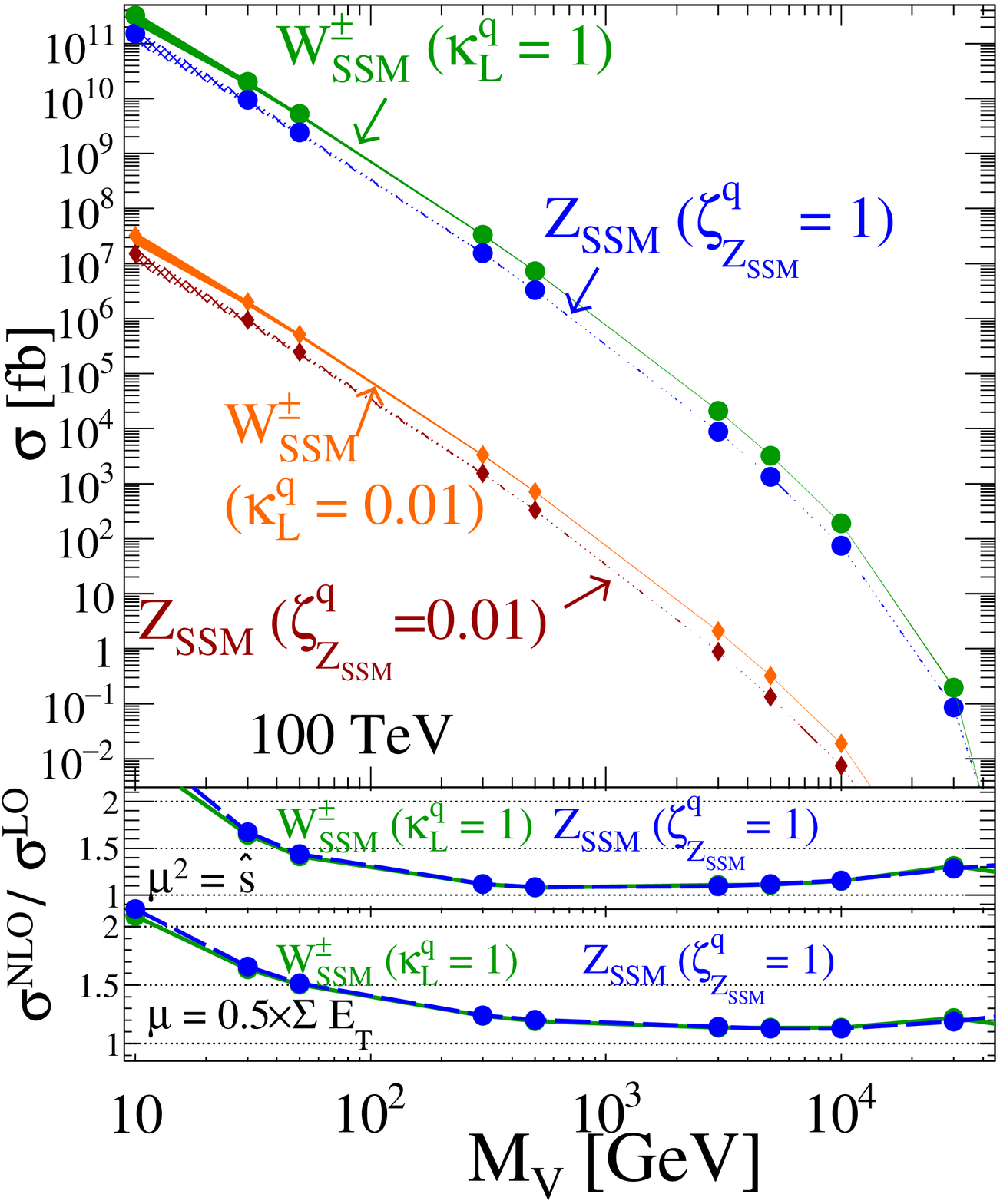}
    \label{fig:xsec100TeV}
  }
  \caption{Total NLO $pp\rightarrow\Wssm$ and $pp\to\Zssm$ production cross
    section at a center-of-mass energy of $\sqrt{s} = $13~TeV (a) and 100~TeV
    (b). The thickness of the curves corresponds to the residual scale
    uncertainty obtained by independently varying the central renormalization
    and factorization scales by a factor of two up and down.}
    \label{fig:xsecTotal}
\end{figure}

As a function of the vector boson mass, we show in Figure~\ref{fig:xsecTotal}
the total inclusive $pp\rightarrow\Wssm$ (solid fill) and $pp\rightarrow\Zssm$
(hatch fill) production rates evaluated at NLO in QCD, assuming the inputs 
listed in Section~\ref{sec:compSetup}.
We set the collision center-of-mass energy to (a) $\sqrt{s}=13$~TeV and (b) 100~TeV,
and use both the benchmark coupling normalizations given in Eq.~\eqref{eq:ssmParamNorm} (circle) 
as well as the much smaller choice (diamond)
\be
 \kappa^{q,\ell}_L=0.01\ , \qquad
 \kappa_R^q=0
 \qquad\text{and}\qquad
 \zeta_{\Zssm}^{f}=0.01 \ .
 \label{eq:smallParamNorm}
\ee
We set as central factorization $(\mu_f)$ and renormalization $(\mu_r)$ scales
half the sum of the transverse energies of all final-state particles,
\be
  \mu_f,\mu_r = \mu_0 = \frac12\sum_{\rm k\in\{final~states\}} E_T^k
  \qquad\text{with}\qquad E_T^k = \sqrt{M_k^2 + p_T^k}\ .
\label{eq:standardscales}\ee
The thickness of each curve in the main panel of Figure~\ref{fig:xsecTotal} 
corresponds to the residual scale uncertainty as
evaluated when varying $\mu_f$ and $\mu_r$ independently 
by a factor of two up and down with respect to the central scale $\mu_0$.
We do not include uncertainties associated with parton distribution functions (PDF).
At 13~TeV (100~TeV), the canonical SSM production rates for a boson
mass lying in the [10~GeV, 5~TeV] ([10~GeV, 30~TeV]) range span approximately
\be\bsp
 \Wssm:&\quad
   \confirm{1.0^{+6.3\%}_{-6.8\%}~-~55\times10^{9}~^{+3.9\%}_{-21\%}~\fb
    \qquad \Big(200\times10^{-3}~^{+3.7\%}_{-4.2\%} ~-~
     320\times10^{9}~^{+16\%}_{-31\%}~\fb\Big)}\ ,\\
 \Zssm:&\quad
   \confirm{0.7^{+5.3\%}_{-5.7\%} ~-~ 25\times10^{9}~^{+3.8\%}_{-21\%}~\fb
    \qquad \Big(86\times10^{-3}~^{+3.2\%}_{-3.7\%}  ~-~
     150\times10^{9}~^{+18\%}_{-32\%}~\fb\Big)}\ ,
\esp\ee
where the largest rates and residual scale uncertainties correspond to the
smallest SSM boson masses.
For the coupling scenario of Eq.~\eqref{eq:smallParamNorm}, 
the cross sections reduce precisely by a factor of $10^{-4}$.
As the same mass scales are probed, the uncertainties for both large and small
SSM coupling scenarios are essentially the same.
For electroweak (EW)- and TeV-scale boson masses, 
the residual scale uncertainties reaches the few-to-several percent level.
However, unlike NNLO contributions, 
threshold resummation effects for $M_{\Wssm/\Zssm}/\sqrt{s}\gtrsim0.3$ 
greatly exceed the NLO uncertainty band~\cite{Mitra:2016kov}.

In the lowest panel of the figures, we show NLO $K$-factor defined as the ratio
\be
  K^{\rm NLO} \equiv \cfrac{\sigma^{\rm NLO}(pp\rightarrow A +X)}
    {\sigma^{\rm LO}(pp\rightarrow A +X)}\ ,
\label{eq:kNLODef} \ee
for the standard scale choice of Eq.~\eqref{eq:standardscales}.
For $M_{V}>\mathcal{O}(10^2-10^3)\GeV$, we observe for both collider
energies  that NLO QCD corrections are, as
expected~\cite{Sullivan:2002jt}, modest, with $K^{\rm NLO}$ remaining below
$\confirm{\sim1.4}$. At smaller masses, NLO corrections are 
large and $K^{\rm NLO} \gtrsim 2$ for $M_{V}\sim 10\GeV$.
In the middle panel of the figures, we evaluate again Eq.~\eqref{eq:kNLODef}
but instead with a central scale choice of the partonic center-of-mass energy,
\begin{equation}
 \mu_0 = \sqrt{\hat s}.
\end{equation}
We observe the same
qualitative dependence of $K^{\rm NLO}$ on the gauge boson mass $M_{V}$, which
suggests that the $K$-factor is mostly independent of the scale choice.
For $M_{V}= 10-50\GeV$, the large $\mathcal{O}(\alpha_s)$ correction 
is interpreted as the dominance  of the $gq\rightarrow V' q'$ channel 
where the final-state quark $p_T$  satisfies $p_T^q > \mu_f$.
For the inclusive NLO $V'$ production cross section, this is a LO-accurate contribution,
and hence suffers from large scale uncertainties.
The largeness of the $\mathcal{O}(\alpha_s)$ corrections
and residual scale uncertainties highlight the importance of computing QCD
corrections for processes sensitive to the deep low-$x$ region of the gluon PDF,
even for EW processes.

In the rest of this work, we focus on the canonical SSM parameterization, 
although our results can be easily generalized to any
framework featuring extra gauge bosons. For arbitrary $\kappa^{q,\ell}_{L,R}$
parameters, the LO $\Wssm$ partial decay widths to fermions are given
by~\cite{Gopalakrishna:2010xm,Han:2012vk,Alwall:2014bza}
\be\bsp
  \Gam{\Wssmp \rightarrow u_i \overline{d}_j'} =&\
    N_c \Big[\kappa_L^{q 2} + \kappa_R^{q 2}\Big]\
    \big|V^{\rm{CKM}}_{ij}\big|^2\ \frac{g^2 \MWprime}{48 \pi}\ ,  \\
  \Gam{\Wssmp \rightarrow t\bar b} =&\
    N_c \Big[\kappa_L^{q 2} + \kappa_R^{q 2}\Big]\
    \big|V^{\rm{CKM}}_{tb}\ \big|^2\ \frac{g^2 \MWprime}{48 \pi}\
    \Big(1-r_t^{\Wssm}\Big)^2 \Big(1+\frac{1}{2}r_t^{\Wssm}\Big)\ , \\
  \Gam{\Wssmp \rightarrow \ell^+ \nu_\ell} =&\
    \Big[\kappa_L^{\ell 2} + \kappa_R^{\ell 2}\Big]\
    \frac{g^2 \MWprime}{48 \pi}\ ,
\esp\ee
with $r_i^{\Wssm} = m_{i}^2/\MWprime^2$. Subsequently, the $\Wssm$ total width
reads, after summing over all final-state flavors,
\be
 \Gamma_{\Wssm} = \Gam{\Wssm \rightarrow u\overline{d}} +
   \Gam{\Wssm \rightarrow c\overline{s}} +
   \Gam{\Wssm \rightarrow t\bar b} +
   3 \Gam{\Wssm \rightarrow \ell \nu_\ell}\ .
\ee
In the canonical SSM where the overall $\Wssm$ coupling strengths are fixed as
in Eq.~\eqref{eq:ssmParamNorm}, the $\Wssm$ branching fraction to a single
lepton species is
\be
 \BR{\Wssm \rightarrow \ell \nu_\ell} \approx
   \frac{g^2 \MWprime / 48 \pi}{g^2 \MWprime(N_c + 1)/16 \pi} =
   \frac{1}{3(N_c + 1)} \approx 8.3\%\ ,
\label{eq:brToLeps} \ee
where the approximation holds in the limit where the $\Wssm$ boson mass
$\MWprime$ is much larger than the top-quark mass $m_t$ and
where the CKM matrix is assumed to be an identity matrix.
In the above expression, the factor of three corresponds to three generations 
with universal couplings, and $(N_c+1)$ to the respective triplet and singlet 
color representations of quarks and leptons.
The small branching fraction shows that $\Wssm$ searches relying 
on leptonic final-state signatures may lack sensitivity in the high-mass region.

\begin{table}
\renewcommand{\arraystretch}{2}
\setlength\tabcolsep{6pt}
\centering
\begin{tabular}{ c || c | c | c | c | c }
  Mass & $30\GeV$ & $300\GeV$ & $500\GeV$ & $3\TeV$ & $5\TeV$ \\
  \hline\hline
  $\Gam{\Wssm}$&$0.760\GeV$ & $8.92\GeV$ & $16.1\GeV$ & $101\GeV$ & $169\GeV$\\
  \hline
  $\Gam{\Zssm}$&$0.802\GeV$ & $8.02\GeV$ & $14.3\GeV$ & $89.6\GeV$ & $149\GeV$\\
\end{tabular}
\caption{LO $\Wssm$ and $\Zssm$ total widths for representative $M_{\Wssm}$ and
   $M_{\Zssm}$ mass values.} \label{tb:benchWidths}
\end{table}

For arbitrary $\zeta_{\Zssm}^{f}$ values, the LO $\Zssm$ partial widths to
fermion-antifermion pairs $f\bar f$ are universally given by
\be
  \Gam{\Zssm\rightarrow f\overline{f}} =
    N_c^f \cfrac{\zeta_{\Zssm}^{f2}g}{12\pi}\ M_{\Zssm}\ \sqrt{1-4r^{\Zssm}_f}
    \Big[g_{A}^{f2}\sqrt{1-4r_f^{\Zssm}} + g_{V}^{f2}(1 + 2r_f^{\Zssm}) \Big]\ ,
\ee
with $r_f^{\Zssm} = m_f^2/M_{\Zssm}^2$, $N_c^f$ being respectively equal to 1
and to 3 for leptonic and quark final states, and where the vector and
axial-vector couplings $g_V^f$ and $g_A^f$ are defined in
Eq.~\eqref{eq:zVACoup}. The branching ratio into a specific leptonic final 
state in the canonical SSM and in the heavy $\Zssm$ limit is about 4\%.

We evaluate in Table~\ref{tb:benchWidths} the total widths of canonical $\Wssm$
and $\Zssm$ bosons for representative masses, using the EW input
parameter values shown below in Eq.~\eqref{eq:smInputs}. In such a
setup, the  bosons are always narrow, so that they could be
discovered by several LHC searches for heavy resonances. 

High-mass dijet resonance search results have constrained, at the 95\% confidence level (CL),
charged SSM boson masses to be above $2.6\TeV$ after analyzing CMS and ATLAS
collision data at a center-of-mass energy of 13~TeV~\cite{ATLAS:2015nsi,%
Khachatryan:2015dcf}, whereas low-mass dijet resonance searches are capable of
excluding leptophobic $\Zssm$ with a mass in the $[350, 500]\GeV$ mass window
for $\zeta_{\Zssm}^q > 0.25-0.26$~\cite{ATLAS:2016bvn}. Extra gauge boson
searches in the leptonic channels currently constrain the neutral canonical $\Zssm$ boson
to have a mass greater than $4.05\TeV$ (in the dileptonic mode)~\cite{%
ATLAS:2016cyf,CMS:2016abv} and the charged $\Wssm$ boson to be heavier than
$4.74\TeV$ (in the single leptonic mode)~\cite{ATLAS:2016ecs,CMS:2015kjy,%
CMS:2016ppa,Khachatryan:2016jww}. 
In terms of couplings, $W'\rightarrow\mu+\MET$ searches at 13 TeV 
imply $W'$ couplings to fermions must obey~\cite{Khachatryan:2016jww}
\begin{equation}
 \kappa^{q,\ell}_L = \frac{g'}{g_{\rm SM}} \lesssim 2.6\times10^{-2} \quad\text{for}\quad \kappa_R^q=0 \quad\text{and}\quad M_{W'} = 300\GeV.
 \label{eq:cmsCoupLimit}
\end{equation}

\subsection{Computational Setup}\label{sec:compSetup}
For concreteness, we consider as a benchmark scenario a SSM model 
with five flavors of massless quarks and a diagonal CKM matrix
$V^{\rm CKM}$ with unit entries. We fix the EW inputs as in
the 2014 Particle Data Group review~\cite{Agashe:2014kda},
\be
 \alpha^{\rm \overline{MS}}(M_Z) = \frac{1}{127.940}\ ,\qquad
 M_{Z} = 91.1876\GeV \qquad\text{and}\qquad
 \sin^{2}_{\rm \overline{MS}}(\theta_{W};M_Z) = 0.23126\ .
\label{eq:smInputs} \ee

Our phenomenological study relies on automated NLO predictions matched to NNLL jet
veto resummation as computed using the \mgLong-SCET framework~\cite{Alwall:2014hca,Becher:2014aya}.
More precisely, within \mg~(v2.5.1), one-loop virtual contributions are numerically evaluated by the
{\sc MadLoop} package~\cite{Hirschi:2011pa} and combined with the real
contributions using the Frixione-Kunszt-Signer (FKS) subtraction
method~\cite{Frixione:1995ms} as implemented in {\sc MadFKS}~\cite{Frederix:2009yq}.
For a jet veto of $\pTVeto$ and a hard process scale $Q$, 
logarithms of the form $\alpha_s^k(\pTVeto)\log^l(Q/\pTVeto)$ with $l\leq2k$ 
are resumed up to the NNLL following the procedure detailed in Section~\ref{sec:jetveto}. 
To generate the necessary UFO model library~\cite{Degrande:2011ua}, 
we design a model file based on the above Lagrangians for the
{\sc FeynRules} program~\cite{Alloul:2013bka} (v2.3.10) that is jointly
used with {\sc NloCT}~\cite{Degrande:2014vpa} and
{\sc FeynArts}~\cite{Hahn:2000kx} (v3.8) for the computation of the 
ultraviolet and $R_2$ counterterms required for numerical one-loop calculations.
Associated UFO files are available publicly from the {\sc FeynRules} model
database~\cite{WprimeZprimeAtNLO}.
Hard scattering events are showered and hadronized using
the {\sc Pythia~8} (PY8) infrastructure~\cite{Sjostrand:2014zea} (v8.212)
and passed to {\sc MadAnalysis~5}~\cite{Conte:2012fm} (v1.4) for
particle-level clustering using the {\sc FastJet} library~\cite{Cacciari:2011ma}
(v3.20) and its implementation of the anti-$k_T$
algorithm~\cite{Cacciari:2008gp}.

Our calculations rely on PDFs and the evaluation of the strong coupling constant
$\alpha_s(\mu_r)$ extracted using the LHAPDF~6 libraries~\cite{Buckley:2014ana} (v6.1.6).
We employ the NNPDF~3.0 NLO PDF sets for LO and NLO calculations,
and the NNLO set for NLO-NNLL calculations~\cite{Ball:2014uwa}. The
factorization and renormalization scales are dynamically set according to Eq.~(\ref{eq:standardscales}).
Following Ref.~\cite{Khachatryan:2016jww}, underlying events are modeled by making use of the PY8 CUETP8M1 tune,
also known as the ``Monash$^*$'' tune~\cite{Skands:2014pea}.

\section{Jet Veto Resummation}
\label{sec:vetoResum}

\subsection{Jet Veto Resummation at Next-to-Next-to-Leading Logarithmic Accuracy with Next-to-Leading Order Matching}
\label{sec:jetveto}
Historically, the first higher order jet veto resummations were carried out in
Refs.~\cite{Stewart:2009yx,Stewart:2010pd,Berger:2010xi}.
In particular, within the SCET framework, jet veto resummation was developed
in parallel in Refs.~\cite{Becher:2012qa,Becher:2013xia,Tackmann:2012bt,Stewart:2013faa}.
To carry out our NNLL jet veto resummation with fixed order NLO
matching, we employ the resummation formalism developed in Refs.~\cite{%
Becher:2012qa,Becher:2013xia} and implemented into \mg~\cite{Becher:2014aya}.
Within SCET~\cite{Bauer:2000yr,Bauer:2001yt,%
Beneke:2002ph,Becher:2014oda}, 
jet veto resummation for the production of a color-singlet, $n$-body final-state system $X$, \textit{i.e.},
\be
 a ~b \to X \quad\text{with}\quad a,b\in\{q,\overline{q},g\},
 \label{eq:bornProcess}
\ee
follows from the existence of the resummed and refactorized fully differential cross
section~\cite{Becher:2012qa,Becher:2013xia},
\be\bsp
 \frac{{\rm d}\sigma^{\rm N^{\it j}LL}(\pTVeto)}
    {{\rm d}y~{\rm d}Q^2~{\rm dPS}_n} = &\ \sum_{a,b=g,q,\overline{q}}
    \Big[\overline{B}_a(\xi_1,\pTVeto)\overline{B}_b(\xi_2,\pTVeto)
       + (1\leftrightarrow2)\Big]\\ &\qquad \times
     \ E_I(Q^2,\pTVeto,\mu_h,\mu,R) \ \mathcal{H}_{ab}(Q^2,\mu_h)
     \  \frac{{\rm d}\hat{\sigma}_{ab}^B(Q^2,\mu)}{{\rm dPS}_n}\ .
\esp\label{eq:diffVetoThm}\ee
Starting from the far right, $\hat{\sigma}_{ab}^B$ is the Born, parton-level
scattering rate for the hard process given in Eq.~\eqref{eq:bornProcess} that 
occurs at a scale $Q$ and with a rapidity $y$. The so-called hard
function $\mathcal{H}_{ab}$ contains the finite virtual corrections to
$\hat{\sigma}^B_{ab}$, and, as non-vanishing loop diagrams factorize in the soft
and collinear limits, is given by the power series
\be
 \mathcal{H}_{ab}(Q^2,\mu_h) = \sum_{k=0}
   \bigg(\frac{\alpha_s(\mu_h)}{4\pi}\bigg)^k \mathcal{H}_{ab}^{(k)}(Q^2,\mu_h)
   = 1 + \frac{\alpha_s(\mu_h)}{4\pi} \mathcal{H}_{ab}^{(1)}(Q^2,\mu_h)
       + \mathcal{O}(\alpha_s^2).
\label{eq:hardFn}\ee
The $\mathcal{H}_{ab}^{(k)}$ coefficients possess logarithms of $(Q/\mu_h)$,
where $\mu_h$ is the scale at which $\mathcal{H}_{ab}^{(k)}$ is regulated,
that can spoil the perturbative convergence of Eq.~\eqref{eq:diffVetoThm} if
$\mu_h\ll Q$. To avoid this, one sets $\mu_h\sim Q$. Details on the efficient
evaluation of the $\mathcal{O}(\alpha_s)$ coefficient $\mathcal{H}_{ab}^{(1)}$
numerically can be found in Refs.~\cite{Hirschi:2011pa,%
Becher:2014aya}. The evolution of the hard process down to scales $\mu, \pTVeto
\ll \mu_h, Q$ is governed by the evolution factor $E_I$,
\be
  E_I(Q^2,\pTVeto,\mu_h,\mu,R) = U_I(Q^2,\mu_h,\mu) \
    e^{-2F_I(\pTVeto,\mu,R)\log\frac{Q}{\pTVeto}}\ e^{2h_I(\pTVeto,\mu)}\ .
\label{eq:evoFn}\ee
The renormalization group evolution function $U_I$ consists of exponentiated
Sudakov form factors and anomalous dimensions. The first exponential in the
above expression is the collinear anomaly that arises from the
breaking of the scale invariance of hadron momenta at the one-loop level in SCET~\cite{Becher:2010tm}. 
At the classical level, the proton momenta are given by $P_i = E_{P_i} (1,0,0,\pm1)$ and the momentum fractions
$\xi_1$ and $\xi_2$ carried by the partons $a$ and $b$ remain unchanged with the
scaling $P_i\rightarrow \tilde{P}_i=\lambda P_i$. One indeed has
\be
  \xi_i \rightarrow \tilde{\xi}_i = \tilde{p}^0_a/\tilde{P}^0_i
    = \lambda p_a^0 / \lambda P^0_i = \xi_i\ .
\label{eq:xiinv}\ee
In particular, simultaneously scaling both proton momenta by $\lambda_i$ and
$\lambda_j = \lambda_i^{-1}$ leaves the hard scale $Q$ unchanged,
\be
 Q \to \tilde{Q}^2 = (4\tilde{p}_a^0\tilde{p}_b^0) =
   (4p_a^0 p_b^0) \lambda_i \lambda_i^{-1} = Q^2 \ .
\label{eq:Qinv}\ee
In the SCET context, while the former invariance of Eq.~\eqref{eq:xiinv} is
broken, the latter one of Eq.~\eqref{eq:Qinv} remains intact.
In the context of perturbative QCD, the collinear anomaly, which arises first at NNLL,
can be understood as the interference between soft virtual corrections and collinear emissions~\cite{Muselli:2017bad}.
The second exponential in Eq.~\eqref{eq:evoFn} is an auxiliary evolution
function that connects the scale $\mu$ to the veto scale $\pTVeto$.
Whereas the indices $a$ and $b$ appearing in $\hat{\sigma}^B_{ab}$,
$\mathcal{H}_{ab}$ and in the beam functions $\overline{B}_a$ and
$\overline{B}_b$ (below) denote specific incoming partons, {\it e.g.}, $a=b=g$
or $a~(b)=u~(\overline{d})$,
the index $I\in\{q,g\}$ in the evolution factor $E_I$ refers to the color representations associated with the
$q\overline{q}$ or $gg$ initial state. This emphasizes the fact that
Eq.~\eqref{eq:diffVetoThm} only holds for color-singlet processes.

Lastly, the beam function for a parton species $a$ in a proton $p$ with a
transverse momentum $p_T^a < \pTVeto$ and carrying a longitudinal momentum
fraction ${\xi_i = E_a/E_{P_i}=e^{\pm y}Q/\sqrt{s}}$ into the hard process is given by
\be
  \overline{B}_a(\xi,\pTVeto) = \sum_{c=g,q,\overline{q}} \int_\xi^1
    \frac{{\rm d}z}{z} ~\overline{I}_{ac}(z,\pTVeto,\mu_f) ~
    f_{c/p}\left(\frac{\xi}{z},\mu_f\right)\ .
\label{eq:beamFn}\ee
The function $f_{c/p}(x,\mu_f)$ denotes the usual transverse-momentum-integrated
density of a parton species $c$ in the proton $p$ carrying a longitudinal
momentum fraction $x=(\xi/z)$ and evolved to a collinear factorization scale
$\mu_f$. The $c\to a$ splitting kernel $\overline{I}_{ac}$ accounts for the
low-$p_T$ (\textit{i.e.}, $p_T<\pTVeto$) collinear splittings of partons that
emerge from $f_{c/p}(x,\mu_f)$ and connects the factorization scale $\mu_f$ to
the veto scale $\pTVeto$. For $\mu_f\sim\pTVeto$, $\overline{I}$ can be expanded
in powers of $\alpha_s$ with coefficients consisting of the Altarelli-Parisi
splitting functions. Moreover, in the $(\pTVeto/Q)\rightarrow0$ limit, emission
recoils can be neglected and the partons $a$ and $b$ in
Eq.~\eqref{eq:bornProcess} remain parallel to their parent protons.

\begin{figure}[t]
  \centering
  \includegraphics[scale=1,width=.96\textwidth]{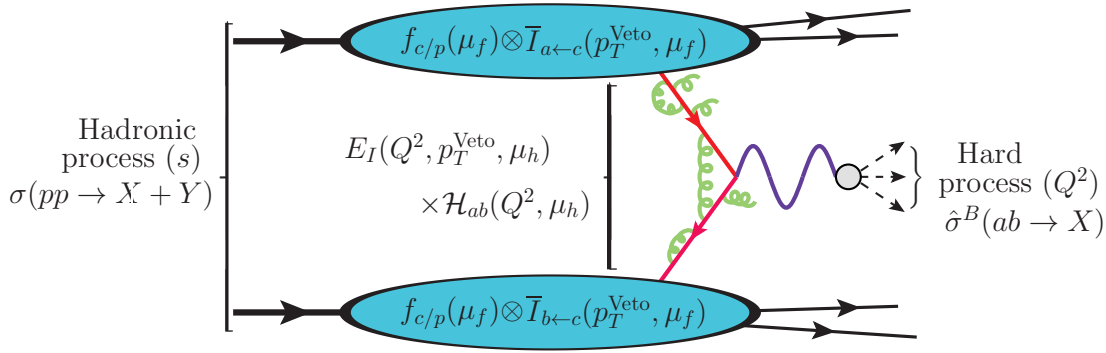}
  \caption{Schematic description of the factorization theorem with jet veto
    resummation in the SCET framework.} \label{fig:factThmSCET}
\end{figure}

The factorization theorem of Eq.~\eqref{eq:diffVetoThm} is illustrated in
Figure~\ref{fig:factThmSCET} and represents the likelihood of the process
$pp\rightarrow X$ to occur when $X$ is accompanied by an arbitrary number of 
QCD emissions possessing a transverse momentum $p_T<\pTVeto$.
It is derived in the $(\pTVeto/Q)\rightarrow0$ limit and hence is valid up to
$(\pTVeto/Q)$ power corrections. For even moderate values of $\pTVeto$,
such contributions are a source of sizable theoretical uncertainties.
These corrections, however, are precisely those that are well-described by
fixed order calculations, which follows from the usual
Collins Soper Sterman (CSS) Collinear Factorization Theorem~\cite{Collins:1985ue}.
The problem can thus be resolved by an appropriate matching procedure
that entails removing the double-counted regions of phase space.
The use of Eq.~\eqref{eq:diffVetoThm}
is necessary due to the breakdown of the CSS formalism in the presence of a jet veto:
A veto prematurely terminates a summation over all collinear, but potentially high-$p_T$, 
emissions that is otherwise necessary for ensuring the cancellation of 
long-range Glauber gluon exchanges~\cite{Zeng:2015iba}.

Matching fixed order and resummed expressions usually involves Taylor expanding
the resummed expression ${\rm d}\sigma^{\rm N^{\it j}LL}(\pTVeto)$ in powers of
$\alpha_s$ to the same accuracy of the fixed order result
${\rm d}\sigma^{\rm N^{\it k}LO}\vert_{p_T<\pTVeto}$. This quantity is then
subtracted from the sum of the fixed order and resummed calculations. For
instance, NNLO matching would require an $\mathcal{O}(\alpha_s^2)$ expansion. In
the SCET framework, the presence of the hard and evolution functions 
marginally complicates the procedure. Extracting these functions, one can rewrite Eq.~\eqref{eq:diffVetoThm} as
\be\bsp
 &\ \frac{{\rm d}\sigma^{\rm N^{\it j}LL}(\pTVeto)}
  {{\rm d}y~{\rm d}Q^2~{\rm dPS}_n} =\sum_{a,b=g,q,\overline{q}}~
    E_I(Q^2,\pTVeto,\mu_h,\mu,R)  ~\times~ \mathcal{H}_{ab}(Q^2,\mu_h)\\
  &\quad \qquad \times \Bigg\{ 
  \bigg[\overline{B}_a(\xi_1,\pTVeto)\overline{B}_b(\xi_2,\pTVeto) +
     (1\leftrightarrow2)\bigg]~
    \cfrac{{\rm d}\hat{\sigma}_{ab}^B(Q^2,\mu)}{{\rm dPS}_n}
    + \Delta \tilde{\sigma}_{ab} \Bigg\}\ ,
\esp\ee
where the $\Delta \tilde{\sigma}_{ab}$ term stands for the $(\pTVeto/Q)$
power corrections with $E_I$ and $\mathcal{H}_{ab}$ factored
out. At the NNLL accuracy, the beam
functions correspond to the Altarelli-Parisi splitting kernels $\overline{I}$
expanded to $\mathcal{O}(\alpha_s)$, which means that the bracketed quantity
represents low-$p_T$ QCD emissions off the Born process up to
$\mathcal{O}(\alpha_s)$. Physically, this is equivalent to the NLO calculation
once a selection on the transverse momentum of the radiated jet of $p_T^j<\pTVeto$ is imposed.
After subtracting the resummed-fixed order overlap, the matched differential jet
veto cross section at the NLO+NNLL accuracy is given, for the generic process
introduced in Eq.~\eqref{eq:bornProcess}, by
\be\bsp
 & \ \cfrac{{\rm d}\sigma^{\rm  NLO+NNLL}(\pTVeto)}{{\rm dPS}_n} =
   \sum_{a,b=g,q,\overline{q}} E_I(Q^2,\pTVeto,\mu_h,\mu,R) ~\times~
   \bigg(1 + \cfrac{\alpha_s(\mu_h)}{4\pi} \mathcal{H}_{ab}^{(1)}(Q^2,\mu_h)
    \bigg) \\
  &\qquad \times \bigg[
    \cfrac{{\rm d}\sigma^{\rm NLO}_{ab}}{{\rm dPS}_{(n+1)}}\bigg|_{p_T<\pTVeto}-
    \cfrac{\alpha_s(\mu)}{4\pi}
    \bigg(\mathcal{H}_{ab}^{(1)}(Q^2,\mu) + E_I^{(1)}(Q^2,\pTVeto,\mu)\bigg)
  \cfrac{{\rm d}\sigma^{\rm LO}_{ab}}{{\rm dPS}_n}
  \bigg]\ .
\esp\label{eq:matchedDiff}\ee
Numerically, the matched result is evaluated over an $(n+1)$-body phase space
domain despite the process in Eq.~\eqref{eq:bornProcess} being an $n$-body
process. The extra emission is however soft by construction, so that each
$(n+1)$-body phase space point is mapped to an $n$-body configuration following
the FKS prescription~\cite{Frixione:1995ms}. With the exception of
${\rm d}\sigma^{\rm NLO}$, all terms are then evaluated within the $n$-body
subspace for the Born process. The relevant analytic expressions for the
ingredients contributing to the matched cross section of
Eq.~\eqref{eq:matchedDiff} can be found in Ref.~\cite{Becher:2014aya} and the
references therein.
{
The NNLL resummation describes the likelihood of process Eq.~(\ref{eq:bornProcess})
being accompanied by up to two soft emissions, 
implying some overlap with the NNLO fixed order calculation.
It is therefore more appropriate to use NNLO PDFs when performing NLO+NNLL computations
as oppose to NLO PDFs, which are needed for NLO computations.
}

\subsection{Non-Perturbative Corrections to Cross Sections with Jet Vetoes}
\begin{figure}
  \centering
  \subfigure[]{
    \includegraphics[scale=1,width=.47\textwidth]{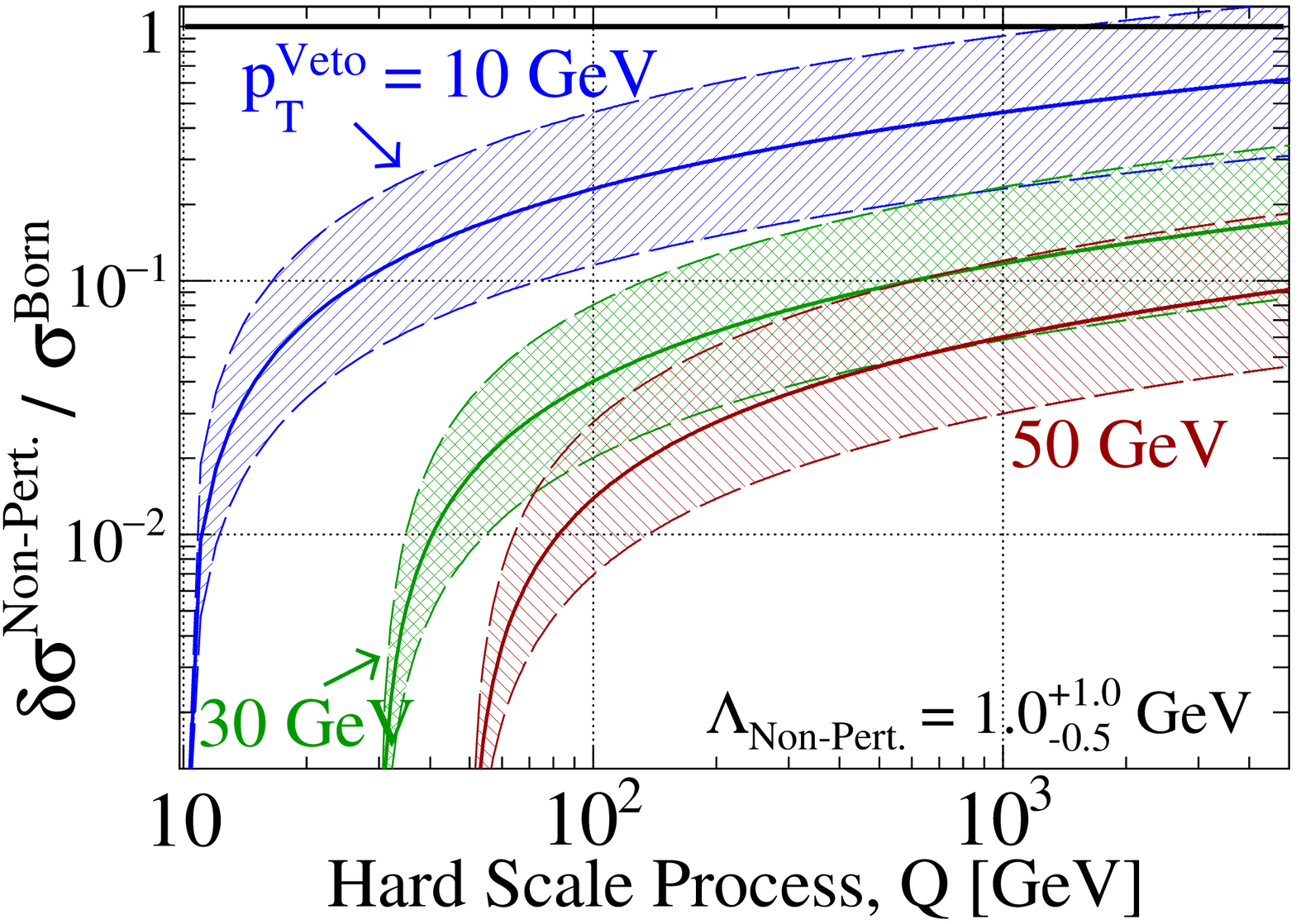}
    \label{fig:nonPertUnc_LoMass}
  }
  \hspace{.25cm}\subfigure[]{
    \includegraphics[scale=1,width=.47\textwidth]{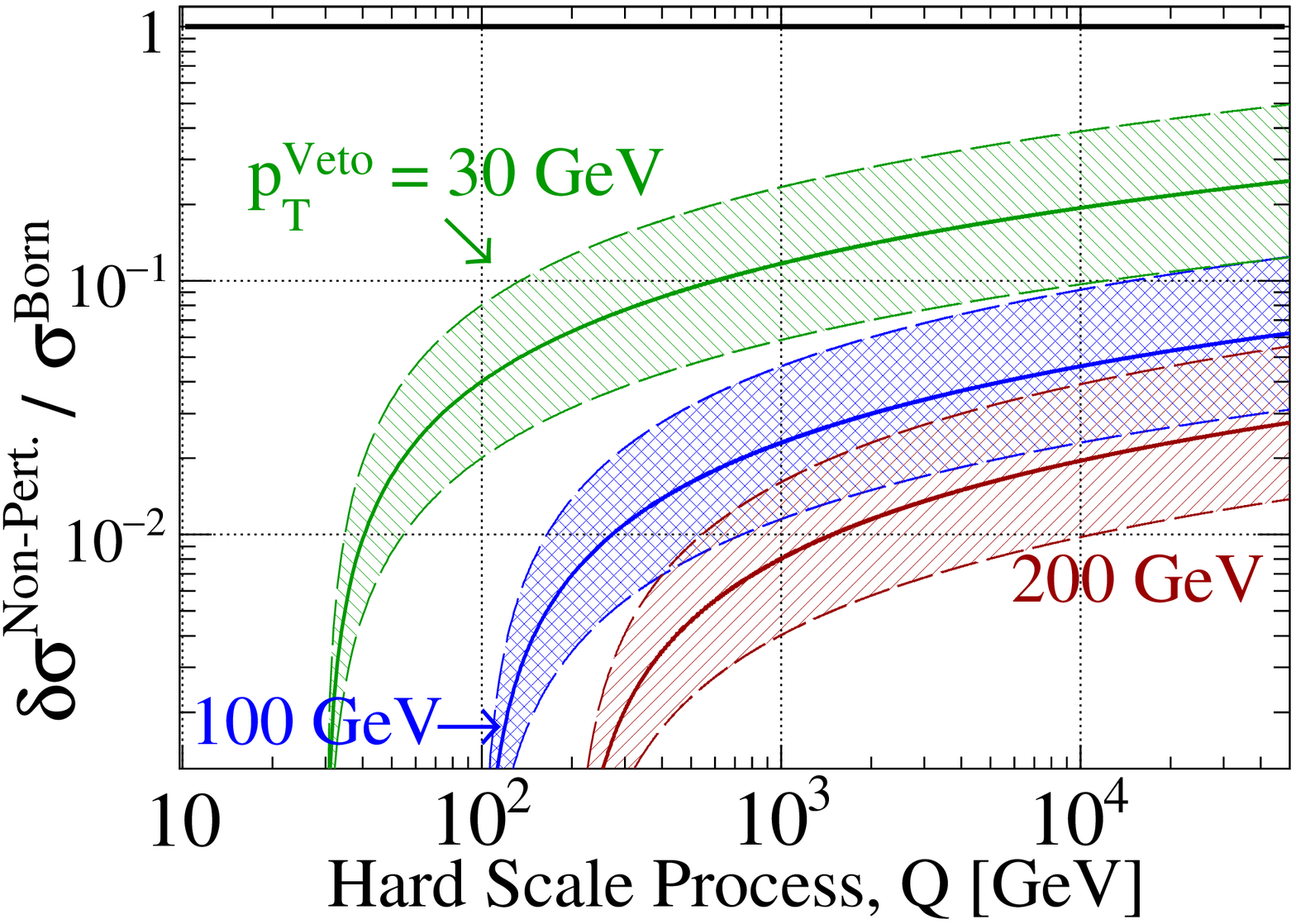}
    \label{fig:nonPertUnc_HiMass}
  }
  \caption{
    Non-perturbative {corrections to jet veto rates}
    arising from the collinear anomaly~\cite{Becher:2013iya,Becher:2014aya} 
    in jet-veto resummed cross sections for low
    (a) and high (b) ranges of the hard process scale $Q$, and for
    representative $\pTVeto$ values. The thickness of the bands reflects the
    variation of the non-perturbative scale $\Lambda_{\rm Non-Pert.}$ as shown
    in Eq.~\eqref{eq:lamvar}.}
  \label{fig:nonPertUnc}
\end{figure}

A consequence of the collinear anomaly in Eq.~(\ref{eq:diffVetoThm}) is the
emergence of logarithmically enhanced non-perturbative corrections that, 
following Ref.~\cite{Becher:2014aya}, are expected to behave as
\begin{equation}
\cfrac{\delta \sigma^{\rm Non-Pert.}}{\sigma^{\rm Born}} \sim \cfrac{\Lambda_{\rm Non-Pert.}}{\pTVeto}\log\left(\cfrac{Q}{\pTVeto}\right)\ ,
\label{eq:nonPertUnc}
\end{equation}
where the energy scale $\Lambda_{\rm Non-Pert.}\sim \mathcal{O}(1-2)\GeV$ is
the scale at which QCD becomes strongly coupled.
Such uncertainties are distinct from non-perturbative corrections to jet observables~\cite{Dasgupta:2007wa},
\textit{e.g.,} shifts in $p_T$ of the hardest jet from out-of-jet emissions of hadrons.
A study of this second class of corrections in the context of jet vetoes is beyond the scope of this report.
However,some of these effects are included due to our use of a modern parton shower
in our NLO+PS-accurate event simulations~\cite{Dasgupta:2007wa}.
For $Q\gg\pTVeto$,
non-perturbative contributions can be sizable. To investigate the impact of
these terms when employing jet vetoes in searches for new color-singlet 
states at hadron colliders, we present the relative magnitude of the
non-perturbative contributions of Eq.~(\ref{eq:nonPertUnc}) as a function of
$Q$ and for representative $\pTVeto$ values in Figure~\ref{fig:nonPertUnc}.
We choose $\Lambda_{\rm Non-Pert.}= \Lambda_{\rm Non-Pert.}^{\rm Default}=1\GeV$
as the central value for the non-perturbative scale, and we include
$\Lambda_{\rm Non-Pert.}$ variation bands  obtained by spanning
\be
   0.5 \times \Lambda_{\rm Non-Pert.}^{\rm Default} ~<~ \Lambda_{\rm Non-Pert.}
     ~<~ 2\times \Lambda_{\rm Non-Pert.}^{\rm Default}\ .
\label{eq:lamvar}\ee
As the $\Lambda_{\rm Non-Pert.}$ dependence in Eq.~\eqref{eq:nonPertUnc} is
linear, these arbitrarily chosen limits induce precisely a variation of a factor
of two up and down around the central value extracted from
Eq.~\eqref{eq:nonPertUnc}.

For  low $\pTVeto$ scales of 10, 30 and 50~GeV, the relative size of the
non-perturbative contribution (with respect to the Born process) respectively
reaches $\delta \sigma^{\rm Non-Pert.}/\sigma^{\rm Born}$ values of about 20\%,
4\% and 2\% for a hard scale of $Q = 100\GeV$. For a larger 
choice of $Q = 1\TeV$, the non-perturbative effects are expected to grow to 50\%,
15\% and 6\%. At an even larger scale of $Q=10\TeV$, the uncertainty originating
from a choice of $\pTVeto=30\GeV$ reaches the 20\% level, whereas it drops to
$\lesssim5\%$ for $\pTVeto \gtrsim 100\GeV$. Those results may suggest that the
linear dependence on the ratio $(\Lambda_{\rm Non-Pert.}/\pTVeto)$ in
Eq.~(\ref{eq:nonPertUnc}) spoils perturbative predictability for overly
aggressive $\pTVeto$ choices when probing mass scales well above the EW
scale. On the other hand, equally aggressive $\pTVeto$ choices for
EW-scale processes give rise to non-perturbative corrections that are
comparable or within the current perturbative and experimental
uncertainties~\cite{Becher:2012qa,Khachatryan:2015sga,Banfi:2015pju}.

For a potential next-generation hadron collider with a center-of-mass energy
well above $13\TeV$, and hence sensitivity to comparably larger hard scales $Q$,
the necessity for choosing $\pTVeto$ at or above the EW scale to avoid
large non-perturbative corrections raises the question of whether or not jet
vetoes are practical for high-mass resonance searches. Standard Model processes,
like $t\overline{t}$ production, dominantly occur near threshold, 
so associated final state momenta scale like the EW scale, and thereby evade such vetoes. 
It may be more advantageous to veto according to a
different metric, such as jet mass. However, it may also be
possible that further investigations into the non-perturbative corrections
induced by the collinear anomaly reveals a milder sensitivity to $\pTVeto$ than
in Eq.~(\ref{eq:nonPertUnc}). 
In particular, one may find for perturbative choices of $\pTVeto$, {\it e.g.,}
$\pTVeto = 30-40\GeV$, where $\alpha_s(\pTVeto) \ll 1$, 
that the non-perturbative contributions turn out to be
negligible when probing multi-TeV hard process scales.

\subsection{Scale Uncertainties in Resummed Jet Veto Rates from Varying Jet Definitions}
The scale dependence of jet veto calculations on the jet definition is sizable but also intuitive: 
For a given hadron collision, a geometrically larger jet will contain more objects and hence will be associated with a larger mass scale. 
This can lead to a larger jet momentum implying that the corresponding event is more likely to be vetoed. 
Furthermore, the lowest order at which the $p_T$ spectrum of any color-singlet system,
which is necessary for calculating jet vetoes in perturbative QCD,
is qualitatively accurate is at NLO for the inclusive process matched to LL$(k_T)$ resummation.
This is also the formal accuracy of NLO+PS calculations used with present day general-purpose event generators.
Similarly, for the veto-resummed calculation, an explicit dependence on the jet radius parameter $R$ of the
$k_T$-style jet algorithms appears first at the two-loop order, {\it i.e.}, at the NNLL level~\cite{Becher:2013xia,Banfi:2012yh}. 
Therefore, predictions provided at NLO+NNLL(Veto) accuracy embeds the lowest order scale dependence on the choice of $R$.

\begin{figure}
  \centering
  \subfigure[]{
    \includegraphics[scale=1,width=.47\textwidth]{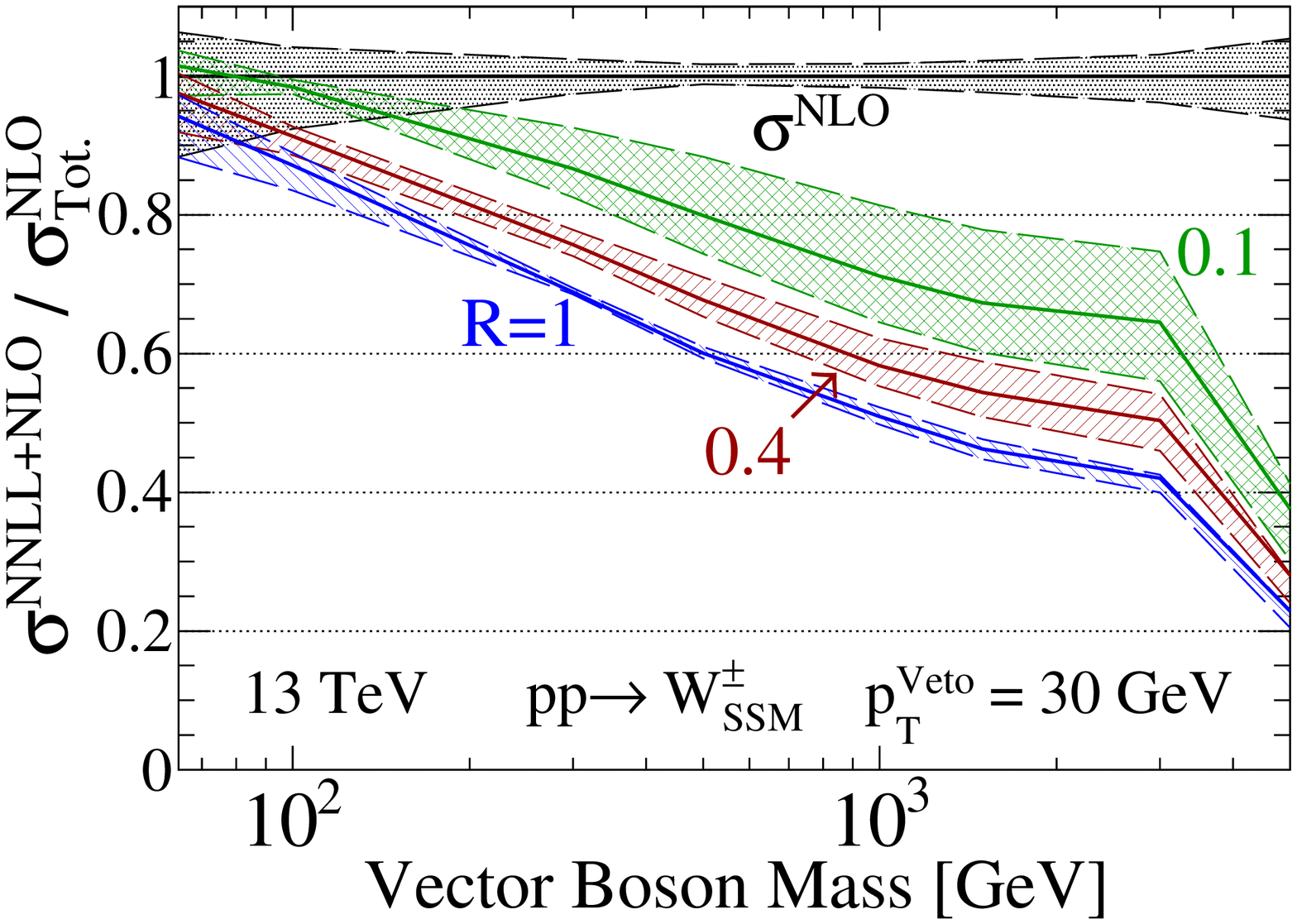}
    \label{fig:jetRadUnc_LoMass}
  }
  \hspace{.25cm}\subfigure[]{
    \includegraphics[scale=1,width=.47\textwidth]{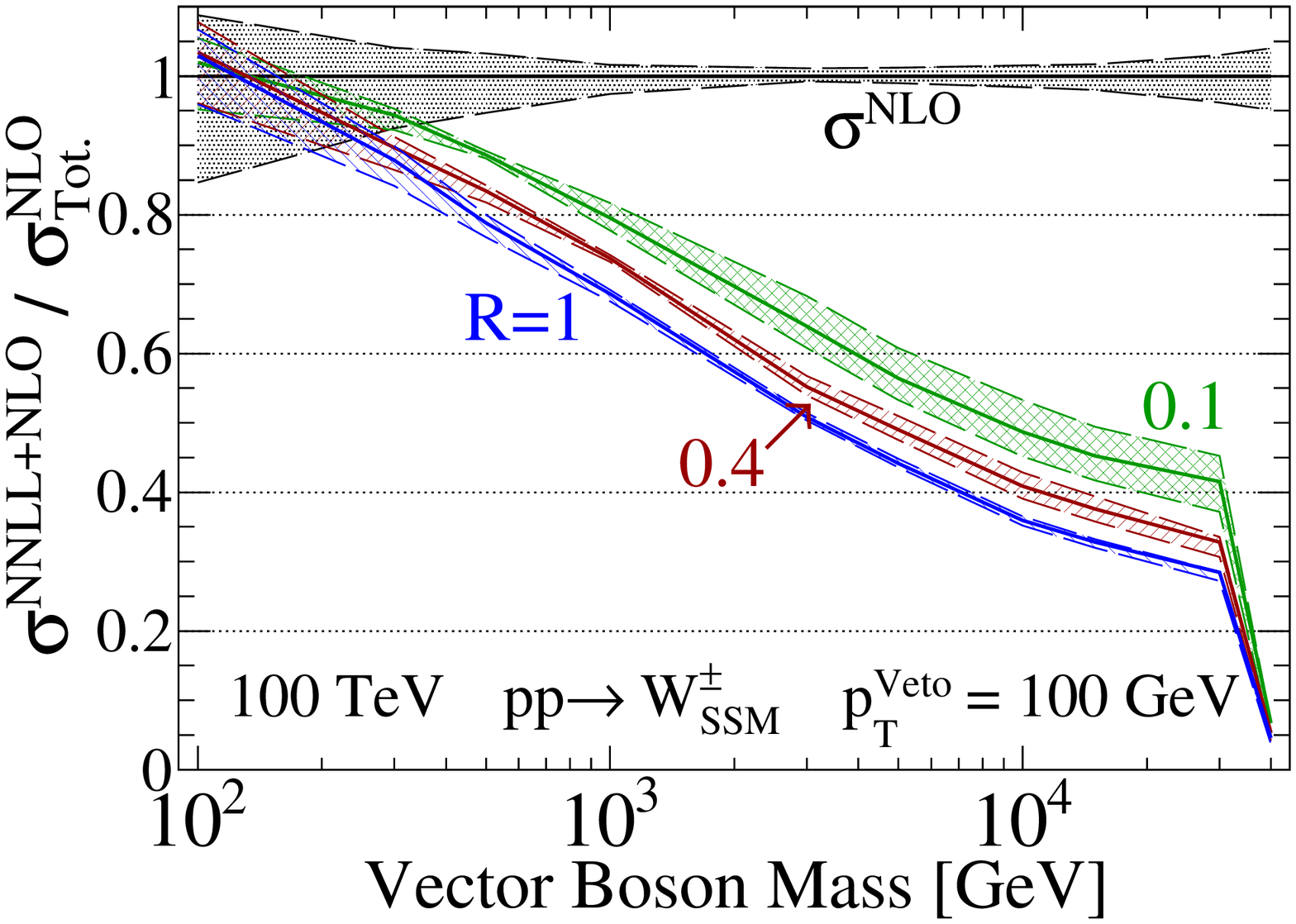}
    \label{fig:jetRadUnc_HiMass}
  }
  \caption{Estimated uncertainty on jet-veto resummed predictions for
    different choices of the jet radius parameter $R$ for $pp$ collisions at a
    center-of-mass energy of 13~TeV (a) and 100~TeV (b). The results are
    presented as a function of the $\Wssm$ boson mass and relatively to the NLO
    total rate $\sigma^{\rm NLO}_{\rm Tot.}$. The uncertainties associated with
    the later are indicated by a black band.}
  \label{fig:jetDefUnc}
\end{figure}

To explore the jet veto scale uncertainty associated with different jet
definitions, we consider the benchmark process
\be
 p p \rightarrow \Wssm 
 \label{eq:vetoUncProcess}
\ee
and focus on two collider energies of $\sqrt{s} = 13$ and 100~TeV.
We restrict ourselves to only investigating the dependence 
on varying $R$ and $\pTVeto$ 
as the veto resummation is identical for all $k_T$-style algorithms at NNLL.
In our choice of representative $R$ values, we are limited by two factors.
First, the factorization theorem of Eq.~(\ref{eq:diffVetoThm}) assumes a jet
radius $R$ satisfying~\cite{Becher:2012qa}
\begin{equation}
 \frac{\pTVeto}{Q} \ll R \ll \log\frac{Q}{\pTVeto}\ ,
\end{equation}
which indicates that $R$ and $\pTVeto$ must obey the relationship
\be
 \pTVeto \ll Q \times e^{-R} \approx Q \Big(1 - R + \frac{R^2}{2}\Big)\ .
\label{eq:rangeveto}\ee
For respectively small, medium and large radius with $R=0.1$, 0.4 and 1, this
translates to $\pTVeto$ scales much smaller than $0.9 Q$, $0.7 Q$ and $0.5 Q$.
For larger $\pTVeto$ scales, matching to the fixed order calculation is necessary
due to a breakdown of the factorization theorem of Eq.~\eqref{eq:diffVetoThm},
as derived in Refs.~\cite{Becher:2012qa,Becher:2013xia}.
The second limitation stems from the logarithmic dependence on $R$ of the
evolution function $E_I$ introduced in Eq.~(\ref{eq:diffVetoThm}).
For very small jet radii, these logarithmic terms are large and need to be
resummed~\cite{Tackmann:2012bt,Dasgupta:2014yra}.
The study of the impact of these
resummed small-$R$ logarithms is beyond the scope of the present
analysis and we refer to Ref.~\cite{Banfi:2015pju} for more information.
For large $R$, the expressions for the anomalous dimensions in Eq.~(\ref{eq:evoFn}) 
break down~\cite{Becher:2014aya}.

In Figure~\ref{fig:jetDefUnc}, we present, as a function of the $\Wssm$ boson
mass, the veto efficiency $\vareps^{\rm NLO+NNLL(Veto)}$ associated with the
process of Eq.~(\ref{eq:vetoUncProcess}),
\be
  \vareps^{\rm NLO+NNLL(Veto)}(\pTVeto) \equiv
    \cfrac{\sigma^{\rm NLO+NNLL(Veto)}(pp\to\Wssm \to\ell\nu_\ell; \pTVeto)}
    {\sigma^{\rm NLO}_{\rm Tot.}(pp\rightarrow \Wssm\to\ell\nu_\ell)}\ ,
\ee
for representative jet radii of $R=0.1$, 0.4 and 1, 
with $\pTVeto = 30~(100)\GeV$, and a collider energy of 13~(100) TeV. 
Shaded bands correspond to the scale uncertainty; PDF uncertainties are omitted.
At both colliders, we observe systematically smaller
efficiencies for larger $R$ values, in agreement with the argument above. For
increasing $\Wssm$ mass, we observe a monotonically decreasing
veto efficiency, which follows from logarithmically-enhanced soft-gluon
emissions that grow as $\alpha_s(p_T^j)\log(Q^2/p_T^{j~2})$ for
$Q\sim M_{\Wssm}$. This tendency for higher mass color singlet processes to
radiate more soft gluons is in addition the basic argument motivating threshold
and recoil resummations.

{As a function of $R$, the associated scale uncertainty shrinks (grows)
with increasing (decreasing) jet radius due to the increasing (decreasing)
inclusiveness of the observable $\varepsilon(\pTVeto)$.}
For $\pTVeto=30\GeV$ at 13~TeV, the uncertainties on the veto
efficiency are of $\delta \varepsilon \sim 10\%$, 5\% and 1\% for $R=0.1$, 0.4
and 1 respectively. For $\pTVeto=100\GeV$ at
100~TeV, they correspondingly drop to $\delta \varepsilon \sim 5\%$, 1\% and
1\%. The fleetingly small uncertainties associated with the $R=1$ jet case are
due to the $R$ dependence in the evolution operator $E_I$ being largely
logarithmic, up to {neglected} power corrections. They are therefore minimized
in the $R\rightarrow1$ limit.

\begin{figure}
  \centering
  \subfigure[]{
    \includegraphics[scale=1,width=.47\textwidth]{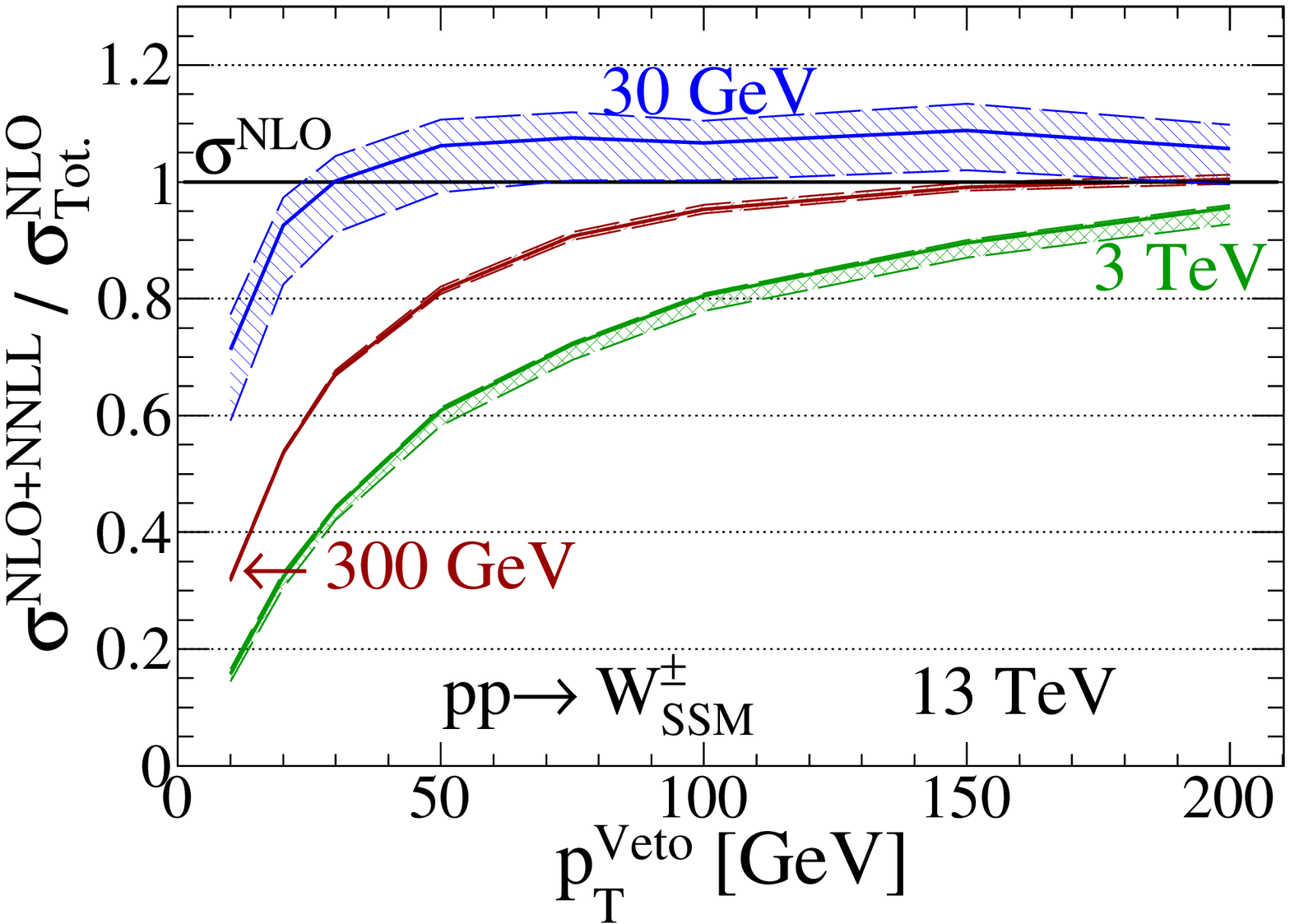}
    \label{fig:veto13TeVNoCuts}
  }
  \hspace{0.25cm}\subfigure[]{
    \includegraphics[scale=1,width=.47\textwidth]{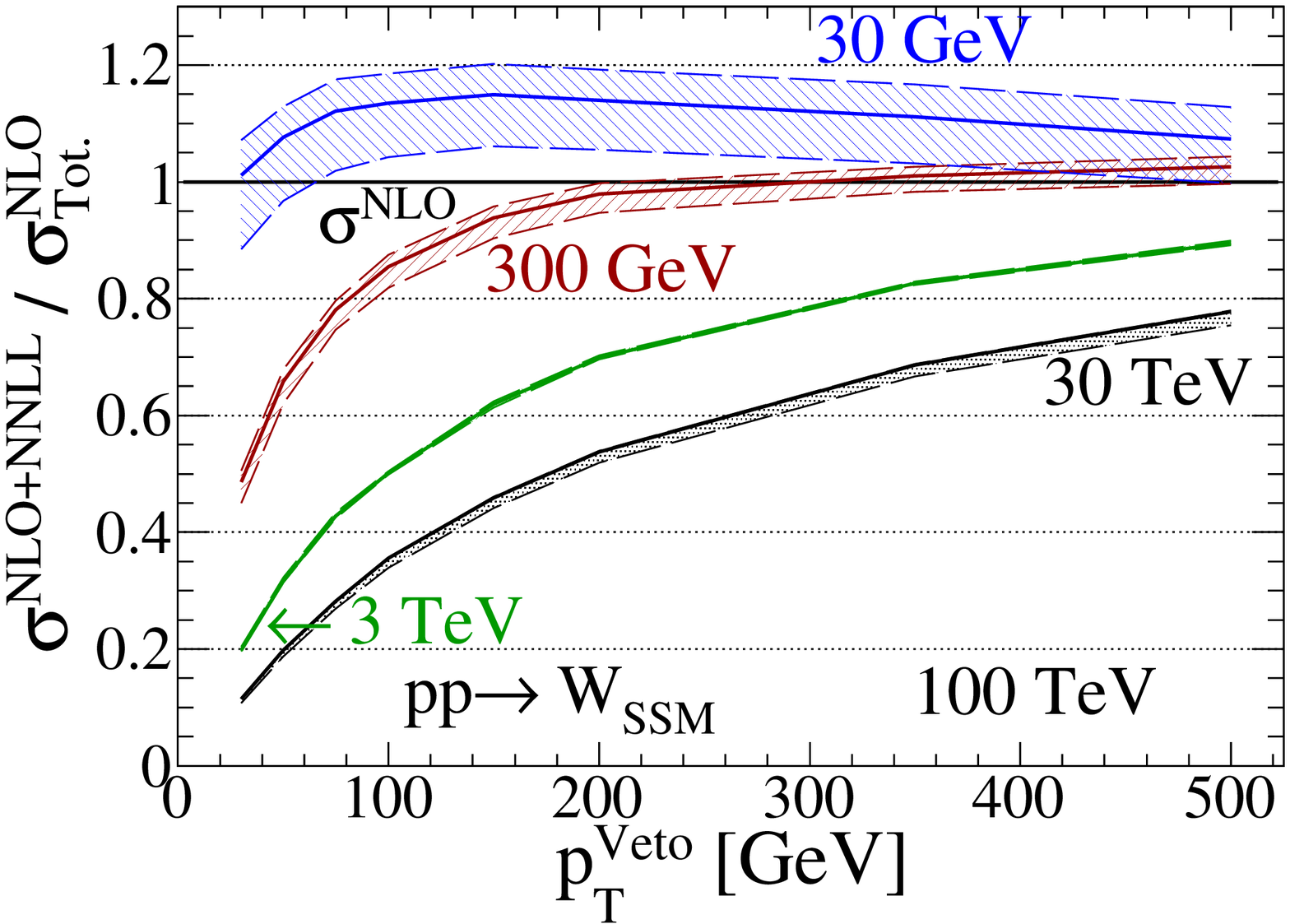}
    \label{fig:veto100TeVNoCuts}
  }
  \caption{Dependence of the resummed jet-veto efficiencies on the veto scale
    $\pTVeto$ for the $pp\to \Wssm \to \mu\nu$ process at a collider energy of
    13~TeV (a) and 100~TeV (b). Basic acceptance selections are included, and we
    consider a representative sample of $\Wssm$ boson masses.}
  \label{fig:pTVetoUnc}
\end{figure}

In Figure~\ref{fig:pTVetoUnc}, we show the scale dependence of the veto
efficiency on the veto scale $\pTVeto$ for representative $\Wssm$ masses
and radius $R=1$ at (a) $\sqrt{s} =13$ and (b) 100 TeV.
The results are consistent with the findings of Figure~\ref{fig:jetDefUnc}:
For a fixed $\pTVeto$,
the production of a heavier $\Wssm$ bosons leads 
to events that are relatively enriched with high-$p_T$ jets, 
which subsequently reduce the veto efficiency. 
As a function of collider energies, increasing $\sqrt{s}$ opens phase space for more jet activity, 
which again implies a smaller veto efficiency. Conversely, for a
fixed $\Wssm$ boson mass, increasing the veto scale increases the veto efficiency
since events are allowed to feature harder jets.
For increasing $\pTVeto$,
$\varepsilon$ converges to 1 and the matched-resummed result converges to the fixed order result, 
as one would expect.
{
However, as the NLO+NNLL result has been computed with NNLO PDFs whereas 
the NLO result with NLO PDFs, there exists a slight mismatch between 
the different central values that is within the (not shown) PDF uncertainties.}

\section{Signal and Background Process Modeling with Jet Vetoes}\label{sec:phenoModeling}
Searches for $W'\rightarrow e/\mu+\MET$ are inundated
with SM Drell-Yan continua and leptonic decays of top quarks. In this section, we
describe our procedure for modeling both the signal and background processes with
jet vetoes. For all processes, we use the computational setup
described at the end of Section~\ref{sec:theory}.

\subsection{$W'$ Production and Decay}\label{sec:wprod}
The benchmark BSM collider signature that we will ultimately simulate
consists of the charged current process
\begin{equation}
 p p ~\rightarrow W' ~\rightarrow \mu ~\nu\ ,
 \label{eq:bsmModelProcess}
\end{equation}
where we consider a final-state muon for the sake of an example.
We use the SSM coupling normalizations of Eq.~(\ref{eq:ssmParamNorm}) and
reinterpret our results for smaller coupling strengths introduced via a multiplicative scaling factor. 
We neglect any interference with the SM $W$ boson due to a severe model
dependence that prevents us from including these effects in a generic way. While
necessary for any SSM-like scenario with a boson mass $M_{W'}$
of the order of the SM $W$ boson mass, little or no such interference is present for
right-handed $W_R$ bosons  in left-right symmetric models or for $W'$ bosons that are 
odd under some discrete symmetry with respect to the SM $W$.

\begin{table}
 \renewcommand{\arraystretch}{1.8}
 \setlength\tabcolsep{6pt}
 \centering
 \begin{tabular}{ c ||  c | c | c || c}
  $M_{W'}$ [GeV] & $\sigma^{\rm NLO}_{\rm Tot.}$ [fb] & 
     $\sigma^{\rm NLO+PS}_{\rm (Veto)}$ [fb]  & $\sigma^{\rm NLO+NNLL}_{\rm (Veto)}$ [fb] &     $K^{\rm NLO+NNLL(Veto)}_{\rm NLO+PS}$ \\
  \hline\hline
 30	 &  $262^{+16\%}_{-25\%}\times10^6$		& $256^{+9.5\%}_{-14\%}\times10^6$	& $296^{+4.5\%}_{-8.2\%}\times10^6$	& $1.16 $ \\
\hline
 50	 &  $68.9^{+9.4\%}_{-17\%}\times10^6$		& $65.6^{+5.7\%}_{-8.7\%}\times10^6$	& $72.8^{+3.4\%}_{-6.4\%}\times10^6$	& $1.11 $ \\
\hline
 300	 &  $289^{+2.1\%}_{-2.8\%}\times10^3$		& $213^{+2.4\%}_{-1.0\%}\times10^3$	& $227^{+0.8\%}_{-0.5\%}\times10^3$	& $1.07 $ \\
\hline
 500	 &  $47.8^{+1.4\%}_{-1.0\%}\times10^3$		& $31.7^{+1.8\%}_{-2.3\%}\times10^3$	& $33.6^{+1.1\%}_{-1.0\%}\times10^3$	& $1.06 $ \\
\hline
 1000	 &  $3.58^{+1.7\%}_{-1.5\%}\times10^3$		& $2.04^{+0.7\%}_{-1.5\%}\times10^3$	& $2.19^{+2.0\%}_{-2.2\%}\times10^3$	& $1.07 $ \\
\hline
 3000	 &  $15.4^{+1.2\%}_{-2.3\%}$			& $7.73^{+0.1\%}_{-1.7\%}$		& $8.06^{+0.6\%}_{-3.4\%}$		& $1.04 $ \\
\hline
 5000	 &  $446^{+1.3\%}_{-1.7\%}\times10^{-3}$ 	& $263^{<0.1\%}_{-0.8\%}\times10^{-3}$	& $258^{+0.7\%}_{-1.8\%}\times10^{-3}$ 	& $0.98 $ \\
\end{tabular}
\caption{Cross sections [fb] for $pp\rightarrow W^{'}\rightarrow\mu\nu_\mu$  at various accuracies
with residual scale uncertainties [\%] (no PDF uncertainties), at the 13 TeV LHC.
  The results are shown for representative   $W'$ boson masses and either without (second column) or with
  (third and fourth columns) a jet veto (for $p_T^{\rm Veto}=\confirm{40\GeV},~R=1$). The
  $K$-factor defined in Eq.~\eqref{eq:kResPSDef} is also indicated (last
  column).}
\label{tb:WSSVetoRates}
\end{table}

We first generate events at the NLO+PS accuracy for the 13 TeV LHC.
At the analysis level, we impose a jet veto 
by rejecting events with $R=1$ jets possessing $p_T^j > \pTVeto=\confirm{40}\GeV$. 
For several representative $W'$  masses, Table~\ref{tb:WSSVetoRates} summarizes
the total inclusive cross section obtained at NLO 
($\sigma^{\rm NLO}_{\rm Tot.}$, second column), as well as
NLO+PS after applying the above jet veto selection ($\sigma^{\rm NLO+PS}$, third column). 
The resummed result
$\sigma^{\rm NLO+NNLL(Veto)}$ is given in the fourth column
of the table and will be used for normalizing the generated NLO+PS events to the
NLO+NNLL(Veto) cross section.
We report residual scale uncertainties [\%]; PDF uncertainties are omitted.
To quantify the impact of this normalization, we
define an appropriate $K$-factor as the ratio of the resummed rate to the NLO+PS
rate once a jet veto event selection is applied,
\begin{equation}
 K^{\rm NLO+NNLL(Veto)}_{\rm NLO+PS}(\pTVeto) 
 \equiv \cfrac{\sigma^{\rm NLO+NNLL(Veto)}(pp\rightarrow W' +X; ~\pTVeto)}{\sigma^{\rm NLO+PS}(pp\rightarrow W' +X; ~\pTVeto)}\ .
 \label{eq:kResPSDef}
\end{equation}
We give, in the last column of Table~\ref{tb:WSSVetoRates}, the corresponding
values for this $K$-factor.
For light $W'$ bosons, the $K$-factors are of the order ${K\gtrsim1.1}$
reduce to ${K\sim1.05}$ for $M_{W'}>100\GeV$,
and drop below this for $M_{W'}>1\TeV$.
In most cases, the PS and resummed results agree within one or two widths of their scale uncertainty bands.
Not shown PDF uncertainties contribute to an additional $\mathcal{O}(1-2)\%$ error.
Below 5 TeV, the $K$-factors are greater than unity,
indicating that the logarithmic corrections in the resummed calculation are positive-definite.
Our $K$-factors are in agreement with the findings of Ref.~\cite{Monni:2014zra} for EW-scale masses
and suggest that the PS and NNLL result converge at much larger mass scales.
The magnitude of the NNLO corrections to the NLO result are known to be comparable in size and negative, 
indicating that the NNLO+NNLL(Veto) result is in agreement with both 
the NLO+PS and NLO+NNLL(Veto) calculations~\cite{Monni:2014zra}.
As the resummed corrections are essentially independent of the hard process,
we expect this behavior to  broadly extend to other color-singlet processes.
Hence, within their given uncertainties, 
both the NLO+PS and NLO+NNLL(veto) calculations give accurate and consistent predictions.
Consequently, jet vetoes applied to color-singlet BSM processes can be reliably modeled 
at the NLO+PS level. This is a main finding of our investigation.

We now briefly comment on whether normalization by Eq.~(\ref{eq:kResPSDef}) is justified at a differential level.
In short, particle kinematics for color-singlet processes in resummed calculations, which possess Born-like kinematics, 
and in NLO+PS calculations, which include recoil from soft and hard radiation, 
are largely the same after applying a jet veto.
This follows from factorization in unbroken gauge theories:
amplitudes containing QCD radiations in the soft/collinear limit factorize into a
product of universal form factors and the (color-connected) Born amplitude. As a
consequence, in this limit, $\mathcal{O}(\alpha_s)$ corrections 
to differential distributions for inclusive DY processes 
reduce to a multiplicative factor applied to the Born cross section. 
Furthermore, this holds analytically for arbitrary $W'/Z'$
couplings and masses~\cite{Ruiz:2015zca}. As the jet veto by definition
removes hard QCD radiations and parton showers are based on collinearly
factorized emissions,
{the kinematics of the two results should therefore 
exhibit differences only of the order of $(\pTVeto/Q)$,
which we assume to be vanishingly small for the validity of the	
jet veto factorization theorem in Eq.~(\ref{eq:diffVetoThm}).}

\begin{figure}
  \centering
  \subfigure[]{
    \includegraphics[scale=1,width=.47\textwidth]{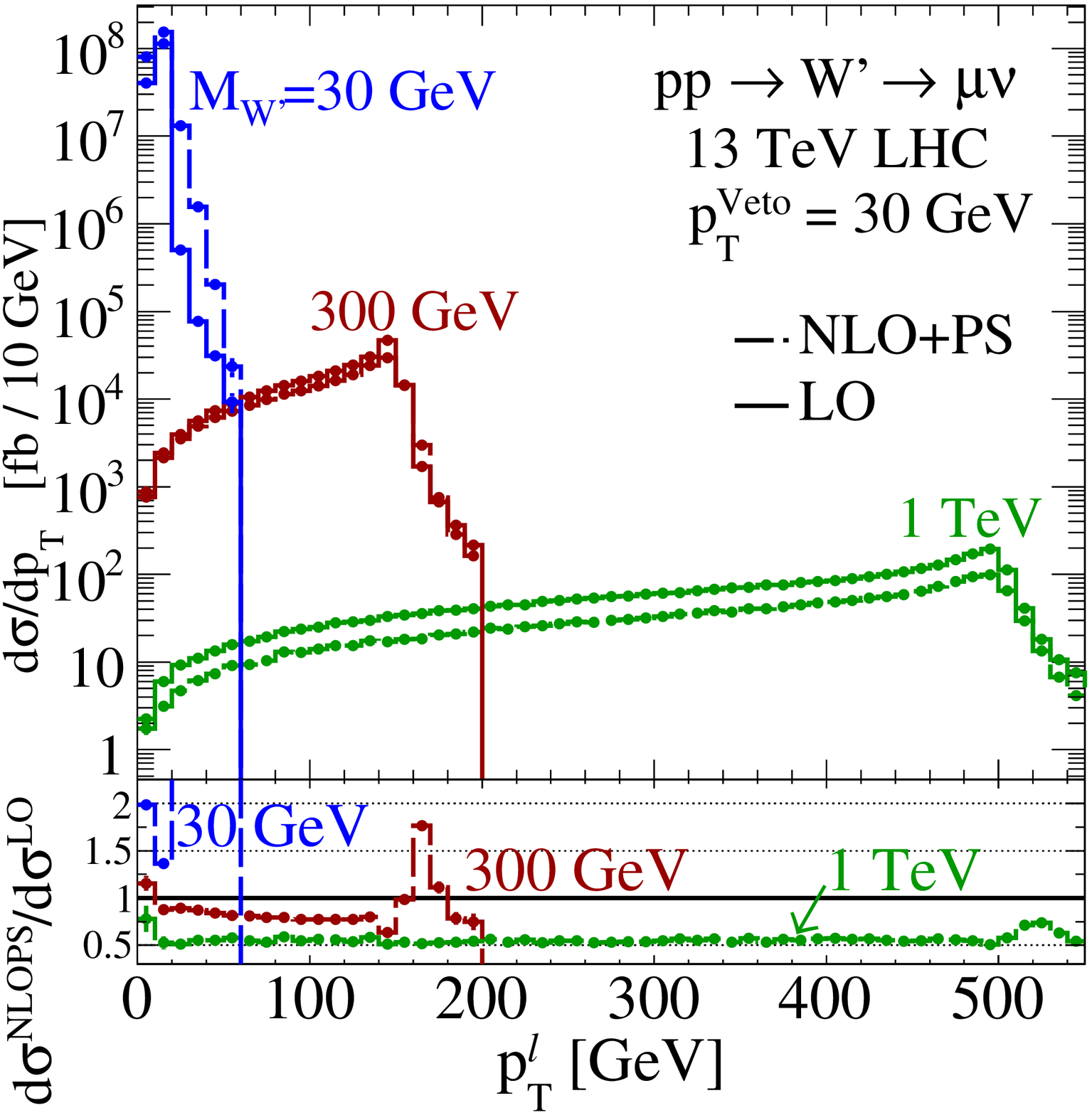}
  }
  \hspace{0.25cm}\subfigure[]{
    \includegraphics[scale=1,width=.47\textwidth]{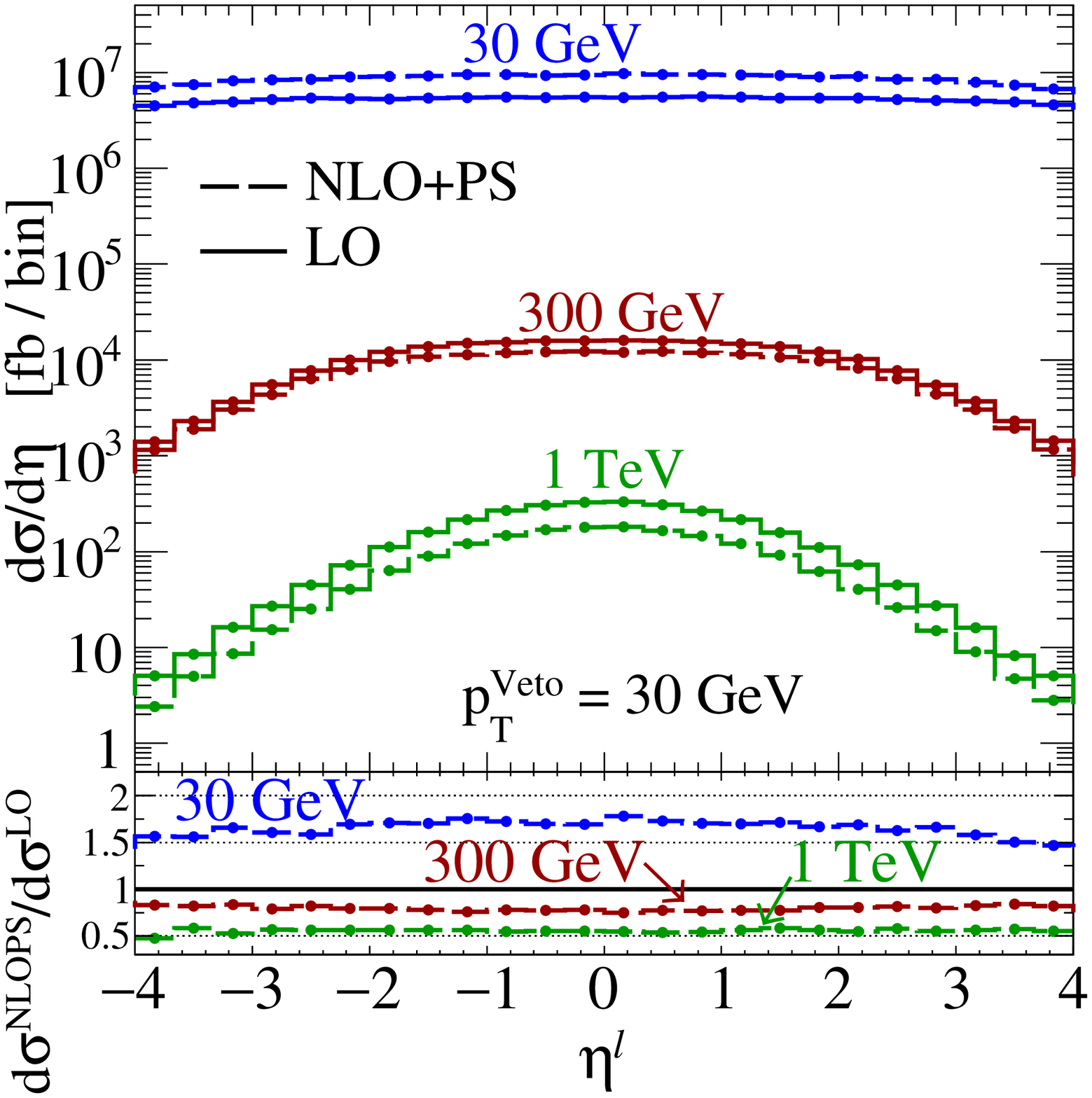}
  }
  \caption{Muon transverse-momentum (a) and pseudorapidity (b) distributions
  for the $pp\rightarrow W'\rightarrow \mu\nu_\mu$ process at a center-of-mass
  energy of 13~TeV. We show results at the LO accuracy (solid) and at the NLO+PS
  accuracy when a jet veto such that $\pTVeto=30\GeV$ is included (dash). We have
  selected a few representative $W'$ boson masses.}
  \label{fig:vetoKin}
\end{figure}

To verify that this holds, we focus on the process in Eq.~(\ref{eq:bsmModelProcess}) and present, 
in Figure~\ref{fig:vetoKin}, the (a) $p_T$ and (b) pseudorapidity $\eta$ distribution of the muon at 13 TeV.
We show results, for
representative $W'$ masses, both at LO (solid) and NLO+PS accuracy with a jet veto of $\pTVeto=30\GeV$ (dash).
At LO, the veto has no impact as no jets are present. In the lower panel of the figure, we
depict the differential NLO+PS $K$-factor for each observable
$\hat{\mathcal{O}}$,
\begin{equation}
 K^{\rm NLO+PS}_{\hat{\mathcal{O}}}(\pTVeto) 
 \equiv \cfrac{{\rm d}\sigma^{\rm NLO+PS}(pp\rightarrow W' +X; 
 ~\pTVeto)/{\rm d}\hat{\mathcal{O}}}{{\rm d}\sigma^{\rm LO}(pp\rightarrow W' +X)/{\rm d}\hat{\mathcal{O}}}\ .
 \label{eq:kNLOPSDef}
\end{equation}
For both distributions, we observe that the bin-by-bin ratios of the LO and
NLO+PS distributions are largely flat when away from resonant regions.
This indicates that the NLO+PS result with a jet veto is dominated by soft gluon radiation, and therefore that the
NLO+PS+$\pTVeto$ kinematics approximate well the jet veto-resummed kinematics.

\subsection{SM Drell-Yan Continua}\label{sec:dyprod}
To model the SM charged and neutral current DY continua, we simulate at
NLO+PS accuracy the processes
\be
  p p \rightarrow W^{(*)} \rightarrow \mu \nu
  \qquad\text{and}\qquad
  p p \rightarrow \gamma^*/Z^{(*)}\rightarrow\mu^+\mu^-\ .
 \label{eq:smDYBkgProcesses}
\ee
For the neutral current channel, we impose a generator-level regulator 
on the dilepton invariant mass $M_{\ell\ell}>10\GeV$. Due to their
color-singlet nature, we treat the DY background much like the signal process,
normalizing the cross sections after including a jet veto by a $K$-factor
such as the one defined in Eq.~\eqref{eq:kResPSDef}.
For the neutral current background, the mass scale is naturally given by the invariant
mass of the dilepton system,  so that for each $M_{W'}$ mass hypothesis, we
derive the $K$-factor by additionally imposing the selection $M_{\ell\ell} > M_{W'}$.
In the charged current case,
constructing the $M_{\ell\nu}$ invariant mass is more subtle due to the 
{(typical)} inability to reconstruct the longitudinal
momentum of final-state neutrinos at the PS level. However, as we discuss in
Section~\ref{sec:sigDef}, we adopt as a discriminating variable sensitive to the
$W'$ mass scale the transverse mass $m_T$ of the lepton-$\MET$ system.
Therefore, for each $M_{W'}$ mass hypothesis, we determine the $K$-factor after
imposing the selection $m_T > M_{W'}$.
Technically, this selection can be implemented in \mg~by identifying
neutrinos as \textit{charged} leptons in the \texttt{SubProcesses/cuts.f} and
\texttt{SubProcesses/setcuts.f} files, and by replacing the $M_{\ell\ell}$
observable by an implementation of the transverse mass $m_T$
in \texttt{SubProcesses/cuts.f}. The relevant selection parameter is thus
{\texttt{mll}}, as in the neutral current case.

\begin{figure}
  \centering
  \subfigure[]{
    \includegraphics[scale=1,width=.47\textwidth]{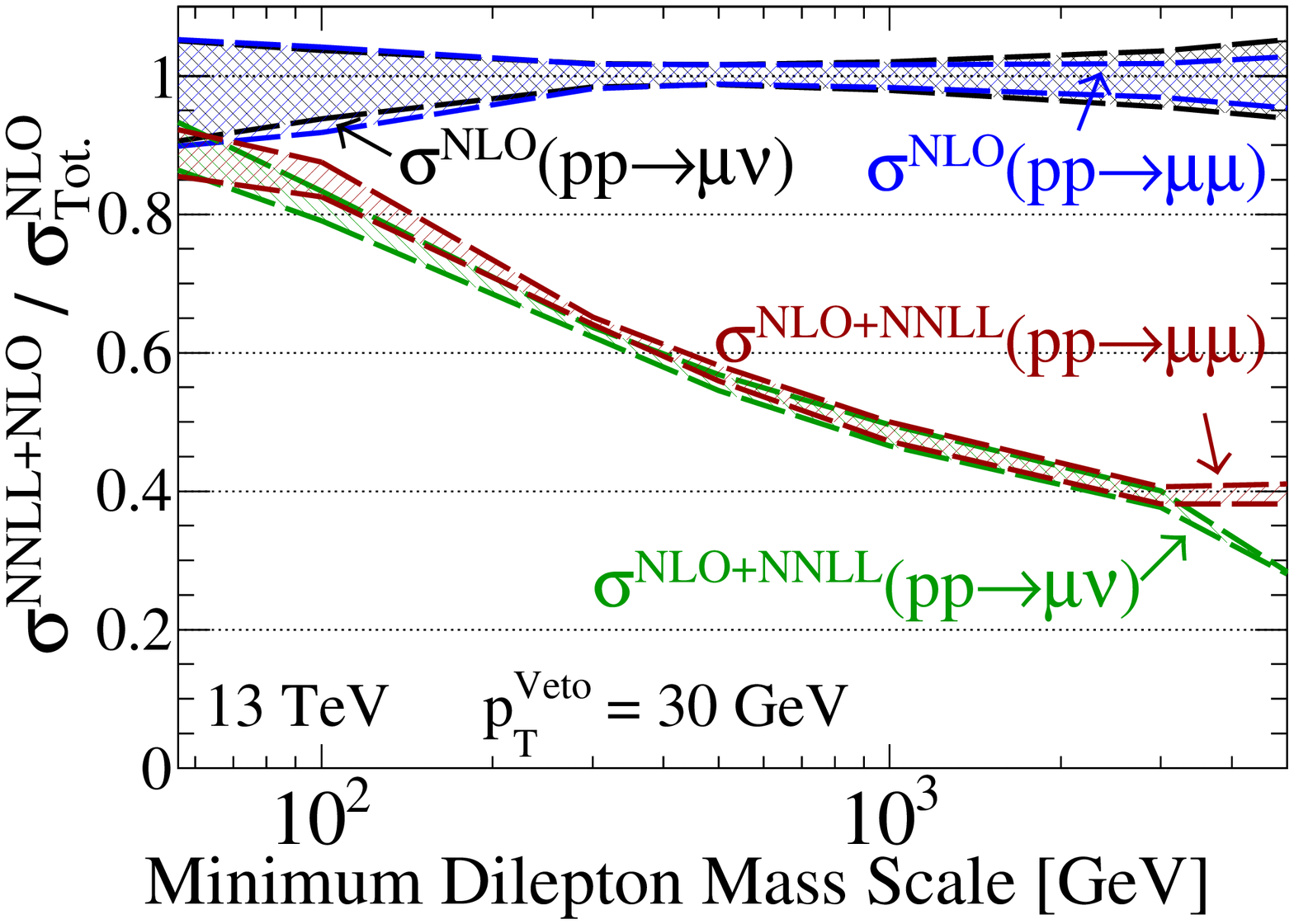}
  }
  \hspace{0.25cm}\subfigure[]{
    \includegraphics[scale=1,width=.47\textwidth]{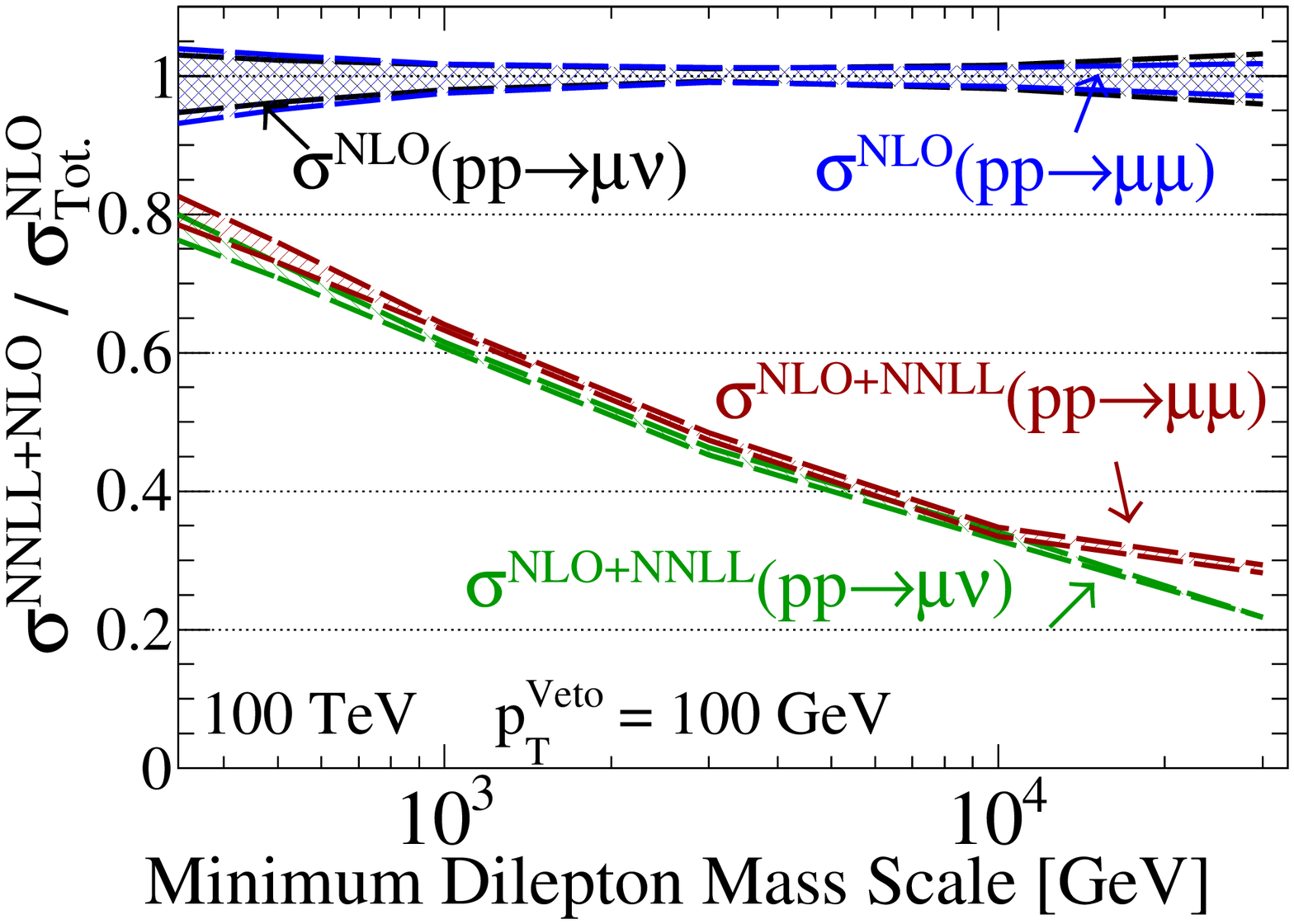}
  }
  \caption{Dependence of the resummed jet-veto efficiencies on the veto scale
    $\pTVeto$ for the neutral current and charged current DY processes at a
    collider energy of
    13~TeV (a) and 100~TeV (b). Basic acceptance selections on the gauge boson
    decay product are included, and we also show the fixed order results.}
  \label{fig:smDYVetoEff}
\end{figure}

In Figure~\ref{fig:smDYVetoEff} we present, as a function of the dilepton mass
scale $M_{\ell\ell}$ and $m_T$ for the neutral and charged current cases
respectively, the veto efficiency for the DY processes given in
Eq.~(\ref{eq:smDYBkgProcesses}) for 
(a) a veto scale of $\pTVeto=30\GeV$ at 13 TeV,
and (b) $\pTVeto=100\GeV$ at 100 TeV.
As anticipated, the impact of the veto becomes more
severe for increasing mass scales, just like the $W'$ case treated in
Section~\ref{sec:wprod}. For both collider and veto setups, we find that
the jet veto efficiencies are independent of the processes and span roughly
\be\bsp
  13\TeV:&\quad
    \confirm{\varepsilon^{\rm NLO+NNLL(Veto)}(\pTVeto=30\GeV) = 90-30\%
    \ \  \text{for}\ \   M_{\ell X} \in [0.050, 5]\TeV,} \\
  100\TeV:&\quad \confirm{\varepsilon^{\rm NLO+NNLL(Veto)}(\pTVeto=100\GeV) =
     80-30\% \ \  \text{for}\ \ M_{\ell X} \in [0.3, 30]\TeV,}
\esp\ee
with a residual scale uncertainty of about \confirm{$\pm1-5\%$}.

In Tables~\ref{tb:smCCDYvetoRates} and \ref{tb:smNCDYvetoRates}, we report, 
for several representative mass scales, 
the inclusive cross sections for the charged current 
and neutral current DY channels, respectively. 
The predictions are given at
NLO ($\sigma^{\rm NLO}_{\rm Tot.}$, second column), 
NLO+PS after applying a jet veto with $\pTVeto=40\GeV$ ($\sigma^{\rm NLO+PS}$, third column),
and after resumming the jet veto effects ($\sigma^{\rm NLO+NNLL(Veto)}$, fourth column). 
The veto $K$-factor defined as in Eq.~\eqref{eq:kResPSDef} is
shown in the sixth column. Overall, we find a good agreement between the
parton showered and resummed predictions given their few-percent-level
uncertainties. For both channels, the $K$-factors are found to span
approximately the $\confirm{1.0-1.1}$ for mass scales above 30 GeV.
Despite the three different scale choices, 
\textit{i.e.,} $M_{W'},~M_{\ell\ell},$ and $m_T$,
we observe the $K$-factors for the signal and background processes 
to be very comparable in size and direction.

\begin{table}
 \renewcommand{\arraystretch}{1.8}
 \setlength\tabcolsep{6pt}
 \centering
 \begin{tabular}{ c ||  c | c | c || c}
  $m_{T}$ [GeV] & $\sigma^{\rm NLO}_{\rm Tot.}$ [fb] &
     $\sigma^{\rm NLO+PS}_{\rm (Veto)}$ [fb] &$\sigma^{\rm NLO+NNLL}_{\rm (Veto)}$ [fb] &      $K^{\rm NLO+NNLL(Veto)}_{\rm NLO+PS}$\\
  \hline\hline
0  & $18.4^{+6.7\%}_{-12\%}\times10^{6}$ 	& $16.5^{+4.1\%}_{-4.7\%}\times10^{6}$ 	& $18.4^{+2.7\%}_{-5.0\%}\times10^{6}$ 	&  $1.12 $\\
   \hline
30 &  $16.3^{+6.3\%}_{-11\%}\times10^{6}$ 	& $14.7^{+4.2\%}_{-4.0\%}\times10^{6}$ 	& $16.1^{+2.6\%}_{-4.8\%}\times10^{6}$ 	&  $1.10 $\\
   \hline
50 & $12.8^{+5.3\%}_{-10\%}\times10^{6}$	& $11.6^{+4.6\%}_{-3.7\%}\times10^{6}$ 	& $12.7^{+2.6\%}_{-4.8\%}\times10^{6}$ 	&  $1.09 $\\
   \hline
100 & $71.2^{+3.7\%}_{-6.2\%}\times10^{3}$	& $61.1^{+2.6\%}_{-1.5\%}\times10^{3}$ 	&  $63.8^{+1.7\%}_{-3.0\%}\times10^{3}$ &  $1.04 $\\
   \hline
300 & $858^{+1.8\%}_{-1.6\%}$ 			& $576^{+1.2\%}_{-1.1\%}$ 		&  $605^{+0.8\%}_{-0.7\%}$ 		&  $1.05 $\\
   \hline
500 & $129^{+1.6\%}_{-1.3\%}$ 			& $78.4^{+3.5\%}_{-0.3\%}$		&  $84.7^{+2.0\%}_{-1.9\%}$ 		&  $1.08 $\\
   \hline
1000 & $7.49^{+2.0\%}_{-2.1\%}$			& $4.01^{+3.5\%}_{-10\%}$		&  $4.229^{+2.4\%}_{-2.9\%}$ 		&  $1.05 $\\
   \hline
3000 & $10.9^{+3.6\%}_{-4.5\%}\times10^{-3}$    & $5.05^{+3.3\%}_{-4.5\%}\times10^{-3}$ 	&  $5.26^{+0.5\%}_{-5.3\%}\times10^{-3}$ & $1.04 $\\
\end{tabular}
\caption{Same as in Table~\ref{tb:WSSVetoRates} but for the SM charged current DY  process.}
\label{tb:smCCDYvetoRates}
\end{table}

\begin{table}
 \renewcommand{\arraystretch}{1.8}
 \setlength\tabcolsep{6pt}
 \centering
 \begin{tabular}{ c ||  c | c | c || c}
  $M_{\mu\mu}$ [GeV] & $\sigma^{\rm NLO}_{\rm Tot.}$ [fb] &
     $\sigma^{\rm NLO+PS}_{\rm (Veto)}$ [fb]  & $\sigma^{\rm NLO+NNLL}_{\rm (Veto)}$ [fb] &     $K^{\rm NLO+NNLL(Veto)}_{\rm NLO+PS}$\\
  \hline\hline
10 & $7.65^{+24\%}_{-32\%}\times10^{6}$ 	& $7.50^{+15\%}_{-21\%}\times10^{6}$ 	& $9.50^{+5.2\%}_{-9.3\%}\times10^{6}$ 	& $1.27 $\\
  \hline
30 & $2.13^{+7.2\%}_{-13\%}\times10^{6}$ 	& $1.91^{+5.4\%}_{-4.6\%}\times10^{6}$ 	& $2.09^{+2.7\%}_{-5.0\%}\times10^{6}$ 	& $1.09 $\\
 \hline
50 & $1.80^{+5.4\%}_{-11\%}\times10^{6}$ 	& $1.61^{+4.7\%}_{-3.5\%}\times10^{6}$ 	& $1.75^{+2.4\%}_{-4.5\%}\times10^{6}$ 	& $1.09 $\\
 \hline
100 & $73.7^{+4.2\%}_{-8.2\%}\times10^{3}$ 	& $62.7^{+2.9\%}_{-3.2\%}\times10^{3}$ 	& $69.2^{+2.2\%}_{-3.9\%}\times10^{3}$ 	& $1.10 $\\
 \hline
300 & $696^{+1.7\%}_{-1.8\%}$ 			& $481^{+1.9\%}_{-1.2\%}$ 		& $510^{+0.7\%}_{-0.5\%}$  		& $1.06 $ \\
 \hline
500 & $111^{+1.6\%}_{-1.2\%}$ 			& $69.6^{+5.3\%}_{-0.4\%}$ 		& $72.8^{+1.5\%}_{-1.4\%}$ 		& $1.05 $ \\
 \hline
1000 & $6.97^{+1.6\%}_{-1.7\%}$			& $3.87^{+0.1\%}_{-0.3\%}$ 		& $3.99^{+2.1\%}_{-2.5\%}$ 		& $1.03 $ \\
\end{tabular}
\caption{Same as in Table~\ref{tb:WSSVetoRates} 
but for the SM neutral current DY  process.}
\label{tb:smNCDYvetoRates}
\end{table}

\subsection{Top Quark Background}\label{sec:tprod}
The top quark background for $W'\rightarrow \ell\nu_\ell$ searches contains both
a top-antitop pair and single top component,
\be
 p p \rightarrow t ~\overline{t} \rightarrow \ell^\pm + ~\MET + X\ , \quad
 p p \rightarrow t ~j \rightarrow \ell^\pm + ~\MET + X\quad\text{and}\quad
 p p \rightarrow t ~W^{*} \rightarrow \ell^\pm + \MET + X\ ,
\label{eq:tprod}\ee
where one or all top quarks decay leptonically for the first two processes, and
where either the top quark or the $W$-boson (or both) proceeds via a leptonic
decay in associated $tW$ production. In the five-flavor scheme, the $s$-channel
$tb$ production mode is included in the $tj$ process definition. We 
ignore additional channels, such as associated $t\overline{t}W/Z/\gamma^*$
production, as they are both coupling suppressed with respect to the 
above processes and possess similar kinematics.

We simulate inclusive $t\overline{t}$ and $tj$ production at  NLO+PS
accuracy. For medium and high $W'$ boson mass, we impose a generator-level
selections on the top quark transverse momentum. 
For the $tW$ channel, we simulate the
$pp\rightarrow t\ell\nu_\ell$ process at LO+PS accuracy.
The difference in
accuracy with respect to the two other processes is necessary to avoid 
double counting of diagrammatic contributions that appear both in the NLO
corrections to the $tW$ process and in the LO contributions to top-antitop
production when using the five-flavor scheme. Whilst a consistent matching of
these two channels at the NLO+PS accuracy has recently been
achieved~\cite{Jezo:2015aia,Jezo:2016ujg}, such a precision is unnecessary for
our purposes. The listed top processes are intrinsically finite at the Born
level and thus do not need regulating selections in the collider signature
definitions. For $t\overline{t}$ and single $t$ production and our scale choices,
we apply $K$-factors of $K = 1.2$
to account for NNLO and threshold resummation corrections beyond
NLO~\cite{Czakon:2013goa,Kidonakis:2010tc,Kidonakis:2012rm}.

Unlike the color-singlet signal and background processes, the top quark channels
inherently give rise to jets that are well-described by fixed order
perturbation theory. At the Born level, the final-state partons that evolve
into jets posses $p_T$ comparable to the hard process scale and are emitted at
wide-angles with respect to the beam axis. Jet vetoes applied to the top quark
background can thus be well-approximated without the need
for resummation {beyond the PS}. Measurements of low jet multiplicities in $t\overline{t}$
production at 8 TeV for instance show good agreement with the theory once both
experimental and theoretical uncertainties are accounted for~\cite{CMS:2016ooc}.
We consequently model the application of jet vetoes to the top background by
simply imposing a $p_T$ selection on the hardest jet present within the detector
fiducial volume after parton showering.

\section{Missing Transverse Energy and Jet Modeling} \label{sec:metJetModeling}
In this section, we discuss the impact of missing energy and jet modeling
in $W'\rightarrow \ell\nu$ searches with jet vetoes.
In particular, we comment on the use of exclusive versus inclusive $\MET$
definitions by ATLAS~\cite{ATLAS:2016ecs} and CMS~\cite{CMS:2015kjy}
respectively, as well as exclusive vetoes (with \textit{e.g.}, anti-$b$-tagging)
versus inclusive vetoes (\textit{i.e.,} which are flavor-summed).
Exclusive $\MET$ is noteworthy as it is potentially a large source of 
systematic uncertainty that has been previously neglected.

\subsection{Exclusive and Inclusive Missing Transverse Momentum}
At 13 TeV, the CMS collaboration uses inclusive $\MET$ in its $W'\rightarrow \ell\MET$ search. 
It is defined in the usual sense as the norm of the transverse momentum imbalance of all visible particles~\cite{CMS:2015kjy},
\begin{equation}
  \MET \equiv \vert \vec{\slashed{p}}_T \vert
  \qquad\text{where}\qquad
   \vec{\slashed{p}}_T = - \sum_{X\in\{\text{visible}\}} \vec{p}_T^X.
\label{eq:metDef}
\end{equation}
Invisible particles are not restricted to light neutrinos, 
but also include ultra-soft and ultra-collinear objects 
as well as anything absorbed by inactive detector material, like screws and nails.
Furthermore, particle identification is based on the particle-flow
technique~\cite{CMS:2009nxa,CMS:2010byl}, which exploits the detector's 
magnetic field and its tracker and electromagnetic calorimeter resolution.
``Blocks'' with known momentum are constructed from tracks and calorimeter
clusters and then identified as particle candidates. 
In a loose sense, the $\MET$ of a CMS event is known before its particle content.

The ATLAS $W'$ boson search of Ref.~\cite{ATLAS:2016ecs} takes a complementary
approach to defining $\MET$ by building the
$\vec{\slashed{p}}_T$ vector from reconstructed objects already satisfying
kinematic and fiducial requirements,
\begin{equation}
  \MET^{\rm Exclusive} \equiv \vert \vec{\slashed{p}}_T^{\rm Exclusive} \vert,
  \qquad\text{where}\qquad
  \vec{\slashed{p}}_T^{\rm Exclusive} = - \sum_{\substack{X\in\{\text{visible~leptons},\\~\text{high-}p_T~\text{jets},~\text{photons}\}}} \vec{p}_T^X.
\label{eq:metDefExcl}
\end{equation}
Specifically, the hadronic contribution includes only $R=0.4$ anti-$k_T$ jets
with $p_T>20\GeV$. Unlike the CMS procedure where the $\MET$
is independent of additional QCD splittings (ignoring pathological regions of
phase space that correspond, for example, to screws and nails), the definition
of Eq.~(\ref{eq:metDefExcl}) by construction does not sum over an arbitrary
number of low energy jets, and hence is an exclusive quantity.
We now investigate the phenomenological consequences of using
Eq.~(\ref{eq:metDefExcl}) as a missing energy definition.

The first impact of using exclusive $\MET$ is the broadening of all $\MET$-based
observables. At the LHC, a $pp$ collision may produce up to a few high-$p_T$
objects but is typically dominated by the high multiplicity production of lower
energy hadrons, \textit{i.e.}, the underlying event and real emissions off the
hard process. While on average particle production is uniform in the transverse
plane, radiation is distributed asymmetrically on an event-by-event basis.
Hence, when clustered with a separation scale of $R$, up to a few
moderate-to-high-$p_T$ jets are balanced transversely by many more low-$p_T$
jets. Excluding the low-$p_T$ jets from the missing energy definition, as done
in Eq.~(\ref{eq:metDefExcl}), thus injects additional missing energy that is
weakly correlated with any real source of $\MET$ that may originate from the
hard process. The issue is exacerbated for smaller jet radii $R' < R$, which
distributes the same momentum from the hard and underlying processes 
over a larger jet multiplicity, thereby decreasing the average jet $p_T$.

\begin{figure}
  \centering
  \subfigure[]{
    \includegraphics[scale=1,width=.47\textwidth]{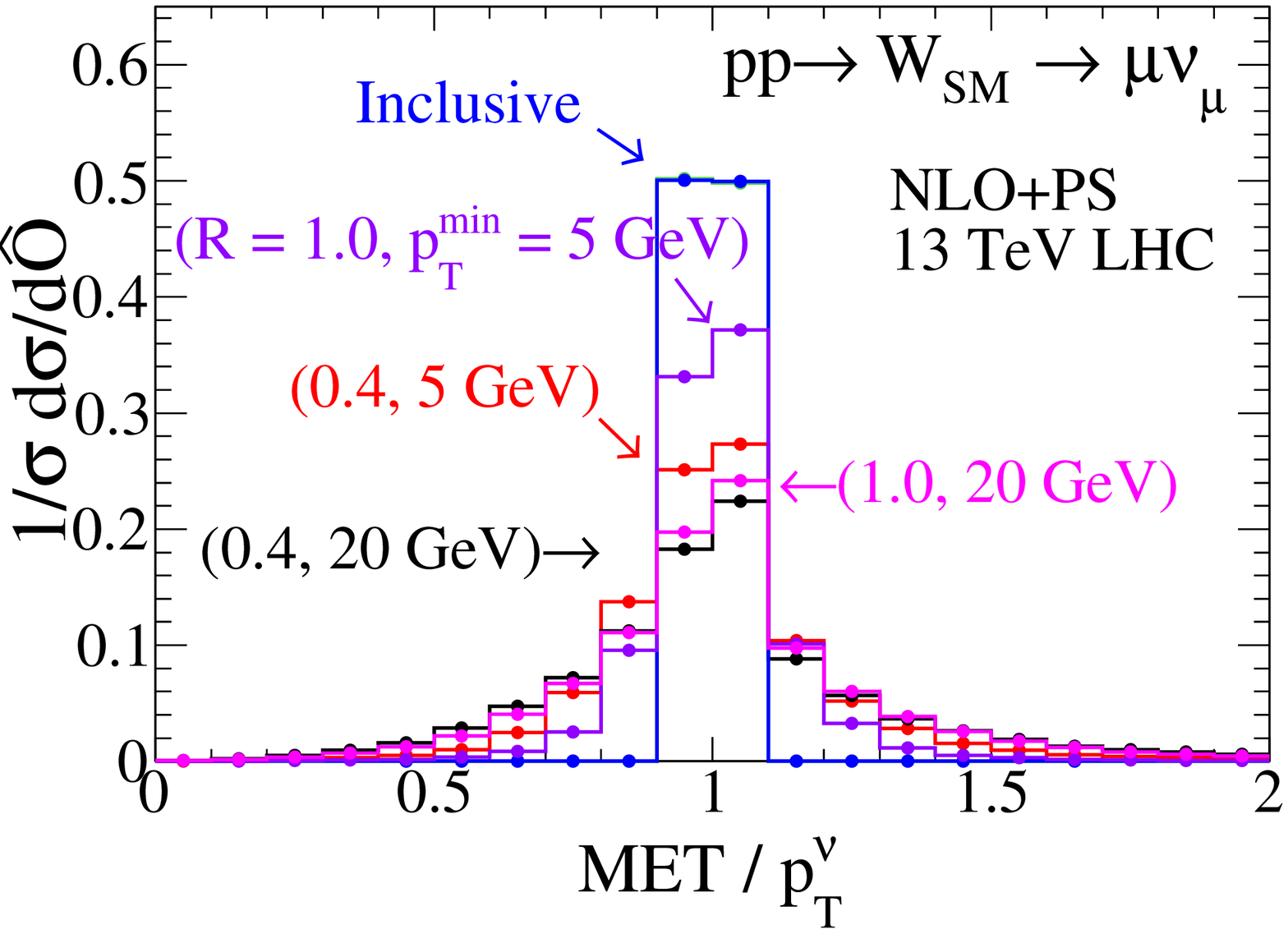}
    \label{fig:vetoPrime_SM_pp_muv_METpTnuRatio}
  }
  \hspace{0.25cm}\subfigure[]{
    \includegraphics[scale=1,width=.47\textwidth]{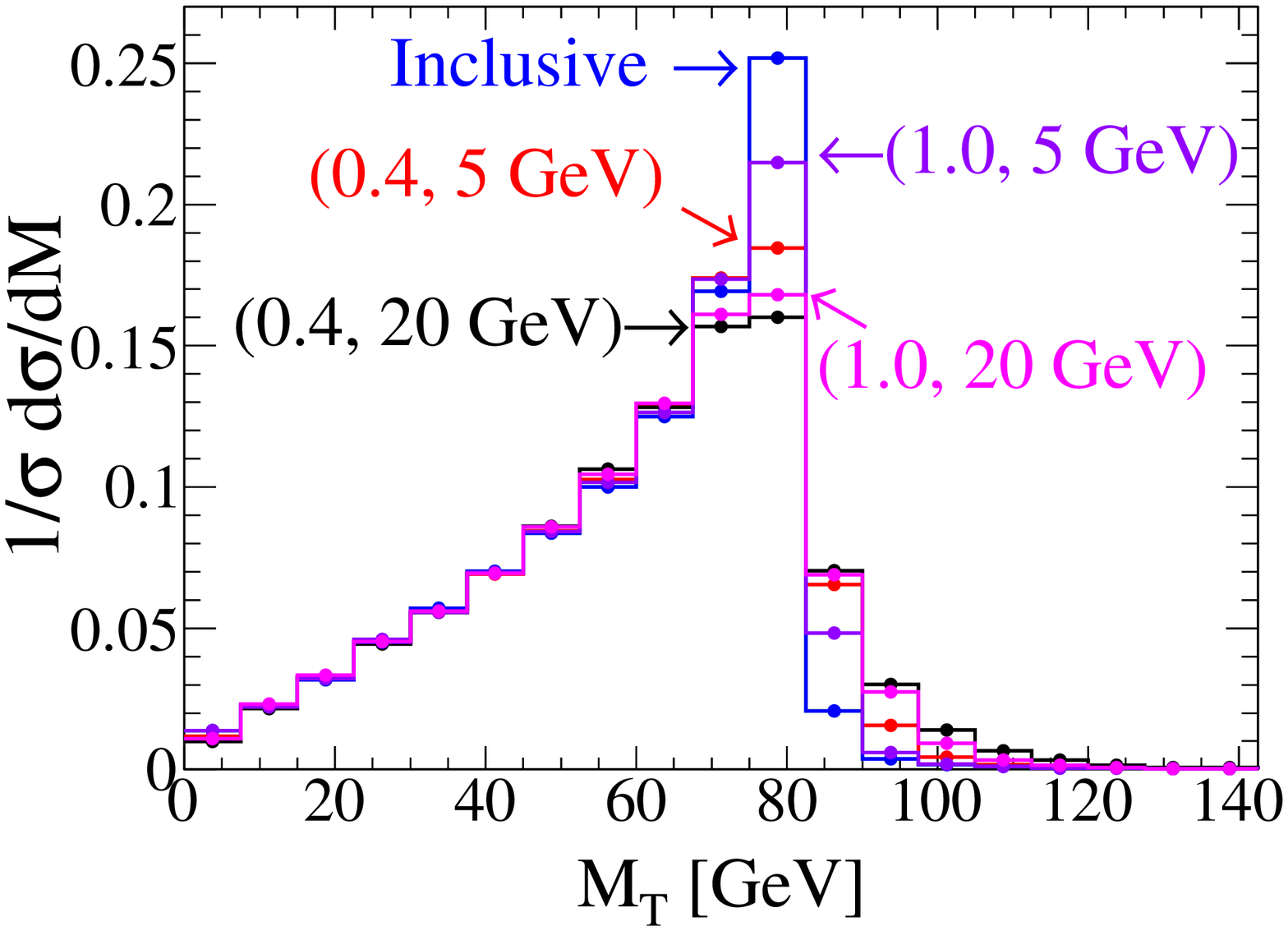}
    \label{fig:vetoPrime_SM_pp_muv_MT}
  }
  \caption{Normalized distributions for the SM 
  process $pp\rightarrow W\rightarrow \mu^\pm \nu_\mu$ 
  in 13 TeV LHC collisions at the NLO+PS accuracy,
  with respect to (a) the ratio of reconstructed $\MET$ 
  to the neutrino $p_T$ and  (b) the transverse mass 
  reconstructed from the $(\ell\slashed{p})$-system,
  using inclusive and exclusive $\MET$ definitions
  and assuming various jet radii and minimum jet $p_T$.}
  \label{fig:metVsJetDef}
\end{figure}

To demonstrate this phenomenon, we consider, at NLO+PS-accuracy, SM $W$ production and decay process
\begin{equation}
 p p \to W \to \mu ~\nu_\mu\ .
\end{equation}
In Figure~\ref{fig:vetoPrime_SM_pp_muv_METpTnuRatio}, we present normalized distributions
for the ratio of the missing transverse energy (both in the inclusive and
exclusive cases) to the transverse momentum of the final-state neutrino. For
both $R=0.4$ and $R=1.0$ jets, the inclusive $\MET$ definition of
Eq.~(\ref{eq:metDef}) describes the light neutrino $p_T$ very well, as one would
expect, with \confirm{more than 99\%} of the distribution being contained within
$\confirm{0.9 < \MET/p_T^\nu < 1.1}$.
On the other hand, requiring jets to satisfy \confirm{$p_T^j >$~5-20\GeV}
reduces this fraction to \confirm{50-80\%}, the rest of the distribution being
smeared evenly around the origin. The broadening is alleviated for
larger jet radii, due to their inherently more inclusive nature. However, the
change is marginal for larger jet $p_T$ requirements.

In Figure~\ref{fig:vetoPrime_SM_pp_muv_MT} is the distribution of the transverse mass reconstructed 
from the $(\mu\slashed{p}_T)$-system, as defined below in Eq.~(\ref{cut:mTDef}). 
For the inclusive case, about \confirm{25\%} of the
distribution is contained in the bin spanning $75\GeV < M_T < 82.5\GeV$. In
contrast, for various exclusive $\MET$ definitions, the peaks drop to about
consisting only of \confirm{16-18\%} of the distribution. Once again a larger
$R$ choice tames the effects due to increase inclusiveness. Consequently, using
exclusive $\MET$ definitions can undermine efforts to search for resonant
structures when using $\MET$-based observables.

A second impact of imposing a $p_T^j > p_T^{\min}$ requirement
in building $\MET$ is the
generation of non-global logarithms (NGLs) of the form 
$\alpha_s\log\left[\MET/(\MET-p_T^{\min})\right]$. NGLs arise when the phase space
associated with virtual corrections of an exclusive observable is
different from the phase space associated with the real corrections~\cite{Dasgupta:2001sh}.
In the inclusive limit, \textit{e.g.,} $p_T^{\min}\rightarrow0$, such logarithms vanish.
Intuitively, NGLs can be understood by imagining a jet that just marginally 
satisfies the $p_T^{\min}$ threshold.  Virtual corrections do not
change kinematics and therefore leave the missing energy unchanged.
However, there exists a corresponding phase space configuration
consisting of a wide-angle emission that brings the initial jet below $p_T^{\min}$. 
Such objects are ignored by the $\MET$ definition of Eq.~(\ref{eq:metDefExcl}) 
and are therefore responsible for inducing shifts in the reconstructed $\MET$ of
order $\Delta \MET\sim \mathcal{O}(p_T^{\min})$. 
This mismatch of the virtual and real phase space configurations can lead to 
potentially large logarithms that would otherwise vanish
for inclusive observables.
Further discussion of resumming such NGLs and the residual scale 
uncertainty are beyond the scope of this study.
Nonetheless, it is clear that the uncertainty associated with the $\MET$ 
reported by ATLAS in Ref.~\cite{ATLAS:2016ecs} is an underestimation.

For our purposes, we employ the inclusive $\MET$ definition of Eq.~(\ref{eq:metDef}).

\subsection{Exclusive and Inclusive Jet Veto}

\begin{figure}
  \centering
  \subfigure[]{
    \includegraphics[scale=1,width=.47\textwidth]{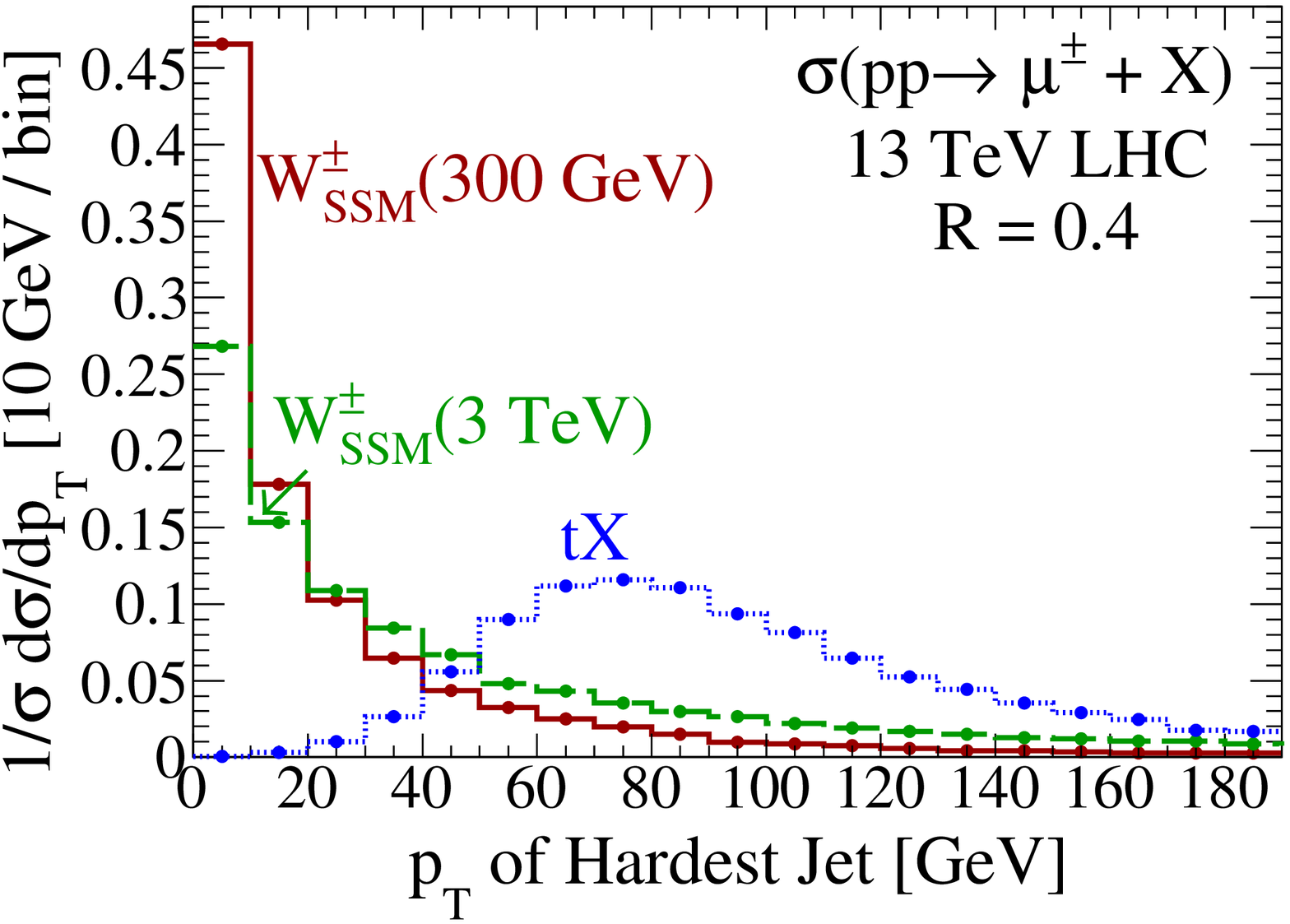}
    \label{fig:JetR0p4_maxJetPt_NoCuts}
  }
  \hspace{0.25cm}\subfigure[]{
    \includegraphics[scale=1,width=.47\textwidth]{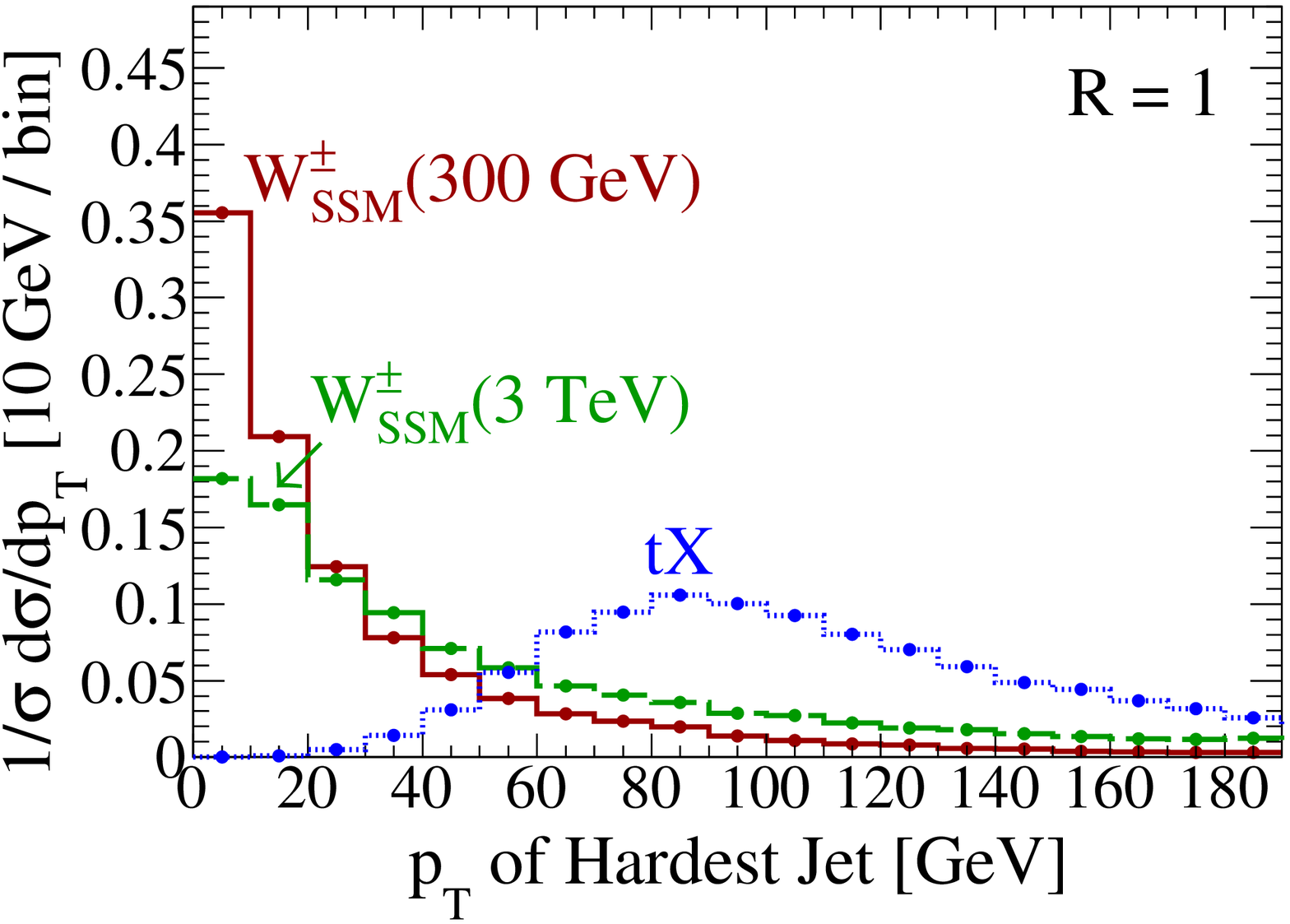}
    \label{fig:JetR1p0_maxJetPt_NoCuts}
  }\\
  \subfigure[]{
    \includegraphics[scale=1,width=.47\textwidth]{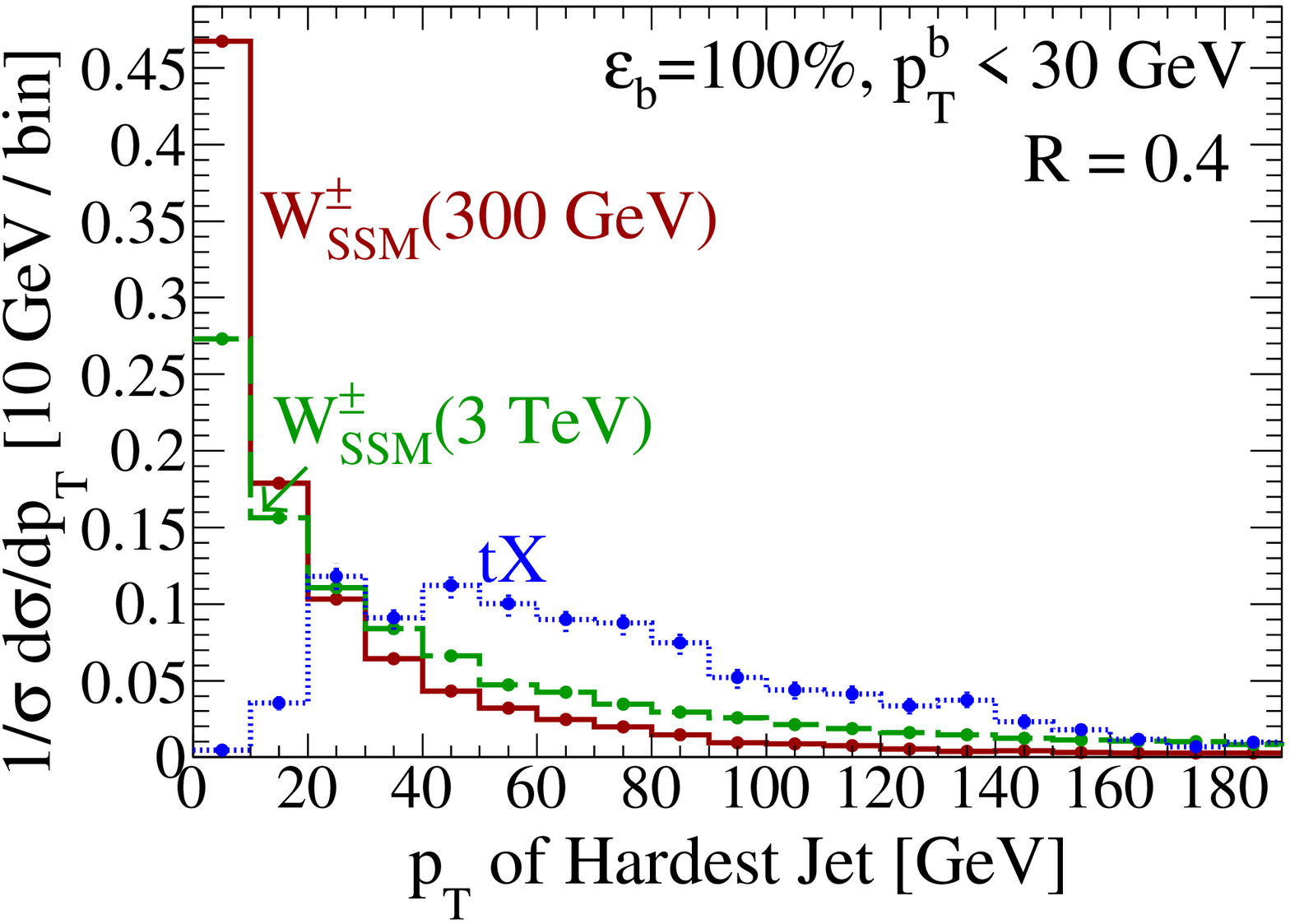}
  \label{fig:JetR0p4_maxJetPt_bjVeto}
  }
  \hspace{0.25cm}\subfigure[]{
  \includegraphics[scale=1,width=.47\textwidth]{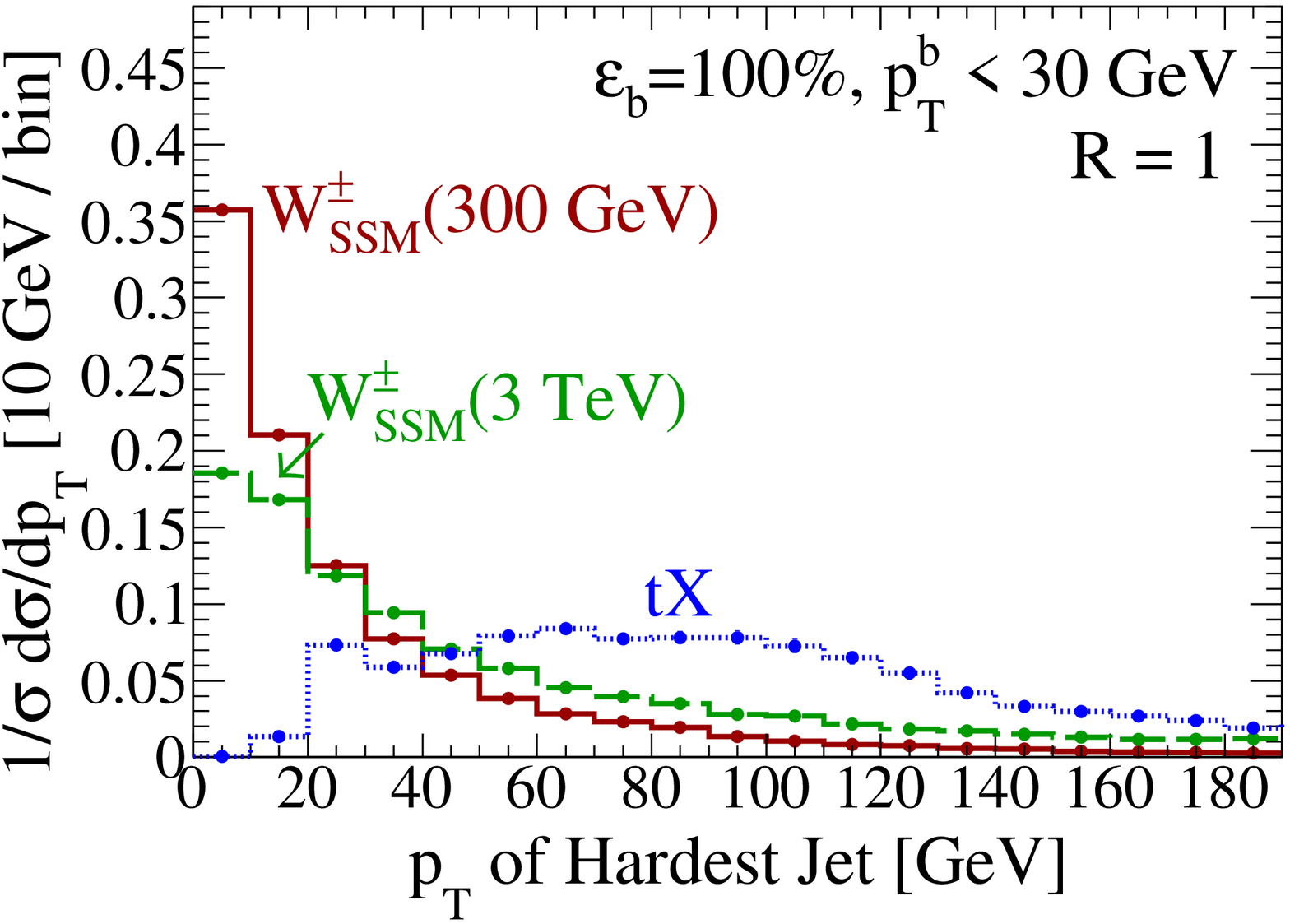}
  \label{fig:JetR1p0_maxJetPt_bjVeto}
  }
  \caption{Normalized $p_T$ distribution of the hardest jet for $W'$ boson and
  single top ($tX$) production in proton-proton collision at a center-of-mass
  energy of 13~TeV. We adopt a jet radius fixed either to $R=0.4$ (a,c) or to
  $R=1$ (b,d), and we either include a $b$-jet veto assuming a tagging/mis-tagging
  efficiency of 100\%/0\%  (c,d) or not (a,b).}
  \label{fig:topBkgKin}
\end{figure}

As for many new physics searches, the leading SM backgrounds for a
$pp\to \Wssm\rightarrow \ell\nu_\ell$ signal include single and pair
production of top quarks shown in Eq.~\eqref{eq:tprod}. As such, it is common
practice to apply anti-$b$-tagging and reject any event featuring a
reconstructed $b$-tagged jet. At the LHC, $b$-tagging methods have typical
identification efficiencies of \confirm{70-80\%} associated with mis-tagging
rates of about \confirm{1.5-10\%}~\cite{Chatrchyan:2012jua}. It is thus
pertinent to determine whether it is valuable to replace exclusive single-flavor
jet vetoes by inclusive flavor-agnostic jet vetoes.
\begin{landscape}
\begin{table}
 \renewcommand{\arraystretch}{1.8}
 \setlength\tabcolsep{6pt}
 \centering
\begin{tabular}{ c | c || c | c || c | c | c }
 & $R$ & $\Wssm(300\GeV)$ & $\Wssm(3\TeV)$ & $t\overline{t}$ & $tj$  &    $t\mu\nu$\\
\hline\hline
  $\sigma_{\rm Tot.}[\text{No.~of~}\mu\geq1]$~[fb]  & N/A  & $24\times10^3$ & $15$ &  $120\times10^3$ & $29\times10^3$ & $6.1\times10^3$\\
\hline\hline \multirow{2}{*}{$b$-Jet Veto [fb]}			  
 & $0.4$ & $24\times10^3$~$(>99\%)$ & $15$~(98\%) & $6.9\times10^3$~(5.6\%) & $5.5\times10^3$~(19\%) & $990$~(16\%) \\
 $\pTVeto=30\GeV,~\vareps^b = 100\%$    
 & $1.0$ & $24\times10^3$~(99\%)  & $15$~(98\%) & $6.1\times10^3$~(5.0\%) & $4.2\times10^3$~(14\%) & $720$~(12\%)\\    
\hline	\multirow{2}{*}{Inclusive Jet Veto [fb]}
		 & $0.4$  &$18\times10^3~(74\%)$ & $7.4~(48\%)$ & $570~(0.5\%)$ & $1.3\times10^3~(5\%)$  		& $160~(3\%)$ \\
$\pTVeto=30\GeV$ & $1.0$  &$16\times10^3~(68\%)$ & $6.3~(41\%)$ & $140~(0.1\%)$ & $670~(2\%)$ 				& $48~(0.8\%)$  \\      
\hline	\multirow{2}{*}{Inclusive Jet Veto [fb]}
		 & $0.4$ & $19\times10^3~(80\%)$ & $8.6~(56\%)$ & $2.1\times10^3~(2\%)$ & $3.5\times10^3~(12\%)$ 	& $400 ~(7\%)$ \\
$\pTVeto=40\GeV$ & $1.0$ & $18\times10^3~(75\%)$ & $7.6~(49\%)$ & $610~(0.5\%)$ & $2.1\times10^3 ~(7\%)$ 		& $150 ~(2\%)$ \\
\hline	\multirow{2}{*}{Inclusive Jet Veto [fb]}
		 & $0.4$ & $20\times10^3~(84\%)$ & $9.5~(62\%)$ & $6.2\times10^3~(5\%)$ & $7.3\times10^3~(25\%)$ 	& $820 ~(14\%)$ \\
$\pTVeto=50\GeV$ & $1.0$ & $20\times10^3~(81\%)$ & $8.6~(56\%)$ & $2.0\times10^3~(2\%)$ & $4.8\times10^3~(16\%)$ 	& $350 ~(6\%)$  \\
\end{tabular}
\caption{Signal and top quark background contributions to the inclusive $pp\to\mu^\pm X$
cross section [fb] and efficiencies $(\%)$ at $\sqrt{s}=13\TeV$ without any jet veto (second line), 
an exclusive $b$-jet veto with $\pTVeto=30\GeV$ and 100\% tagging/0\% mis-tagging efficiency (third line),
and an inclusive jet veto with $\pTVeto=30,~40,~50\GeV$ (fourth, fifth, sixth line) 
for representative jet distance measures $R=0.4,~1$.
}
\label{tb:TopBkgVeto}
\end{table}
\end{landscape}

To assess this, we consider the following $W'$ and top quark production processes at 13~TeV,
\be\bsp
  p p ~\to W'(300\GeV)~\to \mu ~\nu_\mu      &\qquad \text{at~NLO+PS}\ ,\\
  p p ~\to W'(3\TeV)~\to \mu ~\nu_\mu        &\qquad \text{at~NLO+PS}\ ,\\
  p p ~\to t \overline{t} \to \mu^\pm + X    &\qquad \text{at~NLO+PS}\ ,\\
  p p ~\to t j ~\to \mu^\pm + X              &\qquad \text{at~NLO+PS}\ ,\\
  p p ~\to t \mu^\pm \nu_\mu~\to \mu^\pm + X &\qquad \text{at~LO+PS}\ .
\esp\label{eq:bVetoProc5}\ee
As in Section~\ref{sec:tprod}, the differences in the formal accuracies of each
calculation allow us to avoid any possible double counting of diagrams.
We cluster final states into jets as prescribed in Section~\ref{sec:compSetup}.

In Figure~\ref{fig:JetR0p4_maxJetPt_NoCuts}, we show the normalized $p_T$
distributions of the hardest jet from the five processes of
Eq.~\eqref{eq:bVetoProc5}, using a jet radius of $R=0.4$.
For the top backgrounds, a wide plateau can be seen at
$p_T^j \sim m_t(1-M_W^2/m_t^2)/2 \sim 65-70\GeV$, which indicates that
the hardest jet in top production is indeed often a $b$-jet.
For the $W'$ processes, jets are characteristically at a lower $p_T$ value since
their production is entirely occurring via initial-state radiation and they
are thus preferentially soft or collinear to the beam axis.
In Figure~\ref{fig:JetR1p0_maxJetPt_NoCuts}, a jet radius of $R=1.0$ is used and
all distributions are expectedly shifted to higher $p_T$ values.

Assuming an ideal $b$-jet tagging efficiency of $\varepsilon^b = 100\%$ and a
$0\%$ mis-tag rate of a lighter jet as a $b$-jet, we present in
Figure~\ref{fig:JetR0p4_maxJetPt_bjVeto} for $R=0.4$ and
Figure~\ref{fig:JetR1p0_maxJetPt_bjVeto} for $R=1.0$ the same distributions,
but after rejecting all events featuring at least one $b$-jet with a transverse
momentum satisfying $p_T^b > \pTVeto=30\GeV$. Even in this ideal scenario jets
associated with top quarks are still characteristically more energetic, with
$p_T^j \sim M_W/2$, than jets associated with $W'$ production. 
This is related to the sizable single top $tj$ process
which proceeds through a $t$-channel $W$-boson exchange.

We summarize our findings in Table~\ref{tb:TopBkgVeto}.
Here we present the total inclusive production cross section 
(including decays to at least one $\mu$)
for all processes in Eq.~(\ref{eq:bVetoProc5}) (second line),
after applying an exclusive jet veto (third line) 
with $\pTVeto=30\GeV$ and 100\% tagging/0\% mis-tagging efficiency,
as well as after alternatively applying an inclusive veto with 
$\pTVeto = 30,~40,$ and $50\GeV$ (fourth, fifth, sixth lines).
We assume jet radii of $R=0.4$ (above) and $R=1$ (lower).
The corresponding selection efficiencies are shown in parentheses 
and are evaluated with respect to the total rates.
We observe that applying more inclusive jet vetoes,
in terms of both jet radius and flavor composition,
considerably increases the signal-to-noise ratio.
With respect to $b$-jet vetoes, inclusive vetoes of $\pTVeto = 30-40\GeV$
can further suppress top quark production 
by an additional factor of $\confirm{2-50}$
at a modest signal rate cost of $15-50\%$.

\section{Observability of $\Wssm$ with Jet Vetoes at Hadron Colliders} \label{sec:observability}
We now investigate the impact of employing jet vetoes on the discovery potential
of $W'$ bosons in the $pp\rightarrow W'\rightarrow \ell\nu_\ell$ channel at the 13 TeV LHC. 
Simulation of background and signal samples is described in Section~\ref{sec:phenoModeling}.
Our analysis follows, where possible, the 13 TeV SSM $W'$ search methodology employed by the 
CMS collaboration~\cite{CMS:2015kjy,Khachatryan:2016jww}.

We emulate the detector inefficiencies by smearing the momenta of all stable
charged leptons $(\ell=e,\mu)$ and jets reconstructed from the stable hadrons. 
In all cases, the smearing profile is Gaussian~\cite{CMS:2015kjy,Khachatryan:2015mva}, but with different scaling profiles:
For muons, the $p_T$ deviation $(\sigma_{p_T^\mu})$  is parameterized by
\begin{equation}
 \sigma_{p_T^\mu} = a_{p_T}^\mu ~p_T^2\ ,
\end{equation}
where $a_{p_T}^\mu$  is 10\%~TeV$^{-1}$ and 20\%~TeV$^{-1}$ 
for central muons ($\vert\eta^\mu\vert < 0.9$ below 0.9) and forward muons ($\vert\eta^\mu\vert>0.9$),
respectively~\cite{CMS:2015kjy,Khachatryan:2016jww}. The smearing in $p_T$ is then translated into a
change of the energy so that the momentum direction is kept unmodified. 
Unlike its energy scale, the direction of an infinitely energetic stable lepton can still be measured.

Similarly, electron energy uncertainties are parameterized by~\cite{CMS:2015kjy,Khachatryan:2016jww}
\begin{equation}
 \sigma_{E^e} = b_{E}^e ~E \quad\text{with}\quad  b_{E}^e = 4\%\ .
\end{equation}
The difference in parameterizations is due to electron energies being determined
via calorimeters whereas muon momenta are derived from curvature measurements in
a magnetic field. For jets, we follow the 13 TeV CMS $tt+nj$
analysis~\cite{Khachatryan:2015mva} which exploits dedicated energy
calibration and $p_T$ resolution measurements~\cite{Chatrchyan:2011ds}.
Jet energies and $p_T$ are smeared independently according to
\begin{equation}
 \sigma_{\hat{\mathcal{O}}^j} = b_{\hat{\mathcal{O}}}^j \times \hat{\mathcal{O}}
  \qquad\text{for}\qquad  \hat{\mathcal{O}} \in\{E,~p_T\}\ ,
\end{equation}
where the forward (central) coefficient, associated with jet pseudorapidities
satisfying $\vert\eta\vert > 3$ ($\vert\eta\vert < 3$) are fixed to
$b_{E}^j = 3\%$ (5\%) and $b_{p_T}^j = 10\%$ (20\%)~\cite{Chatrchyan:2011ds}.
The change in the jet momentum is translated into a shift in the jet mass,
leaving the jet direction unmodified.

\subsection{Signal Definition and Event Selection}~\label{sec:sigDef}
To test the production of generic $W'$ bosons at colliders, we focus on the
process
\begin{equation}
 p p ~\rightarrow W' ~\rightarrow \mu ~\nu_\mu ~\rightarrow \mu ~+~ \MET\ .
 \label{eq:sigDef}
\end{equation}
The jet veto is agnostic to the lepton flavor; we therefore restrict ourselves to the study of the muon channel for simplicity.
As discussed in Section~\ref{sec:jetveto}, the proposed methodology holds generally for any color-singlet process in hadron collisions, 
including multi-boson and Higgs production.
{
Moreover, the $e+\MET$ mode consists of a multi-jet background~\cite{Khachatryan:2016jww} 
and hence is further enhanced by a jet veto but is otherwise identical to the above channel. 
Applying jet vetoes to the $\tau+\MET$ final state is debatable due to $\tau$ leptons preferential decays to hadrons.}

We identify stable leptons $\ell^\pm$ as hadronically isolated objects for which
the sum of the total hadronic $E_T$ within a distance of
$\Delta R_{\ell X}<0.3$ centered on the the lepton candidate is less than $10\%$
of its $E_T$, \textit{i.e.,}
\begin{equation}
\sum_{X\in\{\text{jets}\}} E_{T}^{X} / E_{T}^{\ell} < 0.1~\quad\text{for}\quad~\Delta R_{\ell X} < 0.3\ .
\end{equation}
We select events containing a single muon candidate meeting 
the following kinematic, fiducial, and leptonic isolation requirements~\cite{Khachatryan:2016jww,CMS:2015kjy}:
\begin{equation}
p_T^{\mu} > \confirm{53\GeV}, ~\quad \vert \eta^\mu\vert < \confirm{2.4}, ~\quad \Delta R_{\mu\ell} > \confirm{0.3} \ .
\label{cut:lepID}
\end{equation}
We reject events with additional isolated electrons and muons satisfying
\be\bsp
 & p_T^{e} >  25\GeV \quad\text{with}\quad 
 \vert \eta^e \vert < 1.444 \quad\text{or}\quad 
 1.566 < \vert \eta^e \vert < 2.5 \ , \\
 & p_T^{\mu} >  35\GeV \quad\text{with}\quad 
 \vert \eta^\mu \vert < 2.4\ .
\esp\label{cut:lepVeto}\ee

Following the results of Sections ~\ref{sec:phenoModeling} and ~\ref{sec:metJetModeling}, we cluster stable
hadrons into jets according to the anti-$k_T$ algorithm~\cite{Cacciari:2008gp} with a separation scale of $R=1$. 
We base our jet veto on the efficiencies of Table~\ref{tb:TopBkgVeto} and reject any 
event with one or more jets whose properties satisfy
\begin{equation}
 p_T^j > \pTVeto = 40\GeV \quad\text{and}\quad \vert \eta^j \vert < 4.7\ .
\label{cut:jetVetoDef}
\end{equation}
We subscribe to the CMS inclusive $\MET$ definition~\cite{CMS:2009nxa,CMS:2010byl}, as given in
Eq.~(\ref{eq:metDef}), and sum over all charged leptons (including non-isolated
objects) as well as all clustered hadronic activity satisfying
$p_T^{\rm Had.} > 0.1\GeV$ and $\vert \eta^{\rm Had.} \vert < 4.7$.

The following selection is then performed to enhance the signal-over-noise $S/B$
ratio~\cite{CMS:2015kjy},
\begin{equation}
  \vert \Delta \phi(\vec{p}^\ell, \vec{\slashed{p}})\vert > 2.5
  \qquad\text{and}\qquad
  0.4 < p_T^\ell / \MET < 1.5 \ ,
\label{cut:cmsWprimeCuts}
\end{equation}
where we respectively constrain the azimuthal separation between the selected
muon and the missing momentum and the ratio of the lepton transverse momentum
to the missing transverse energy. As longitudinal momenta of light neutrinos
cannot be generically inferred in hadron collisions, the transverse mass $(M_T)$
of the $(\ell\MET)$-system,
\begin{equation}
 M_{T} = \sqrt{2 ~p_T^\ell ~\MET~ [1 - \cos\Delta \phi(\vec{p}^\ell,\not\!\!\vec{p})]} \ ,
\label{cut:mTDef}
\end{equation}
is eventually used as a discriminating variable.
$K$-factors accounting for QCD corrections beyond NLO are applied according to the 
prescriptions given in Section~\ref{sec:phenoModeling}.

\subsection{Extended Discovery Potential and Sensitivity at 13 TeV}~\label{sec:discovery}
\begin{figure}
  \centering
  \subfigure[]{
    \includegraphics[scale=1,width=.47\textwidth]{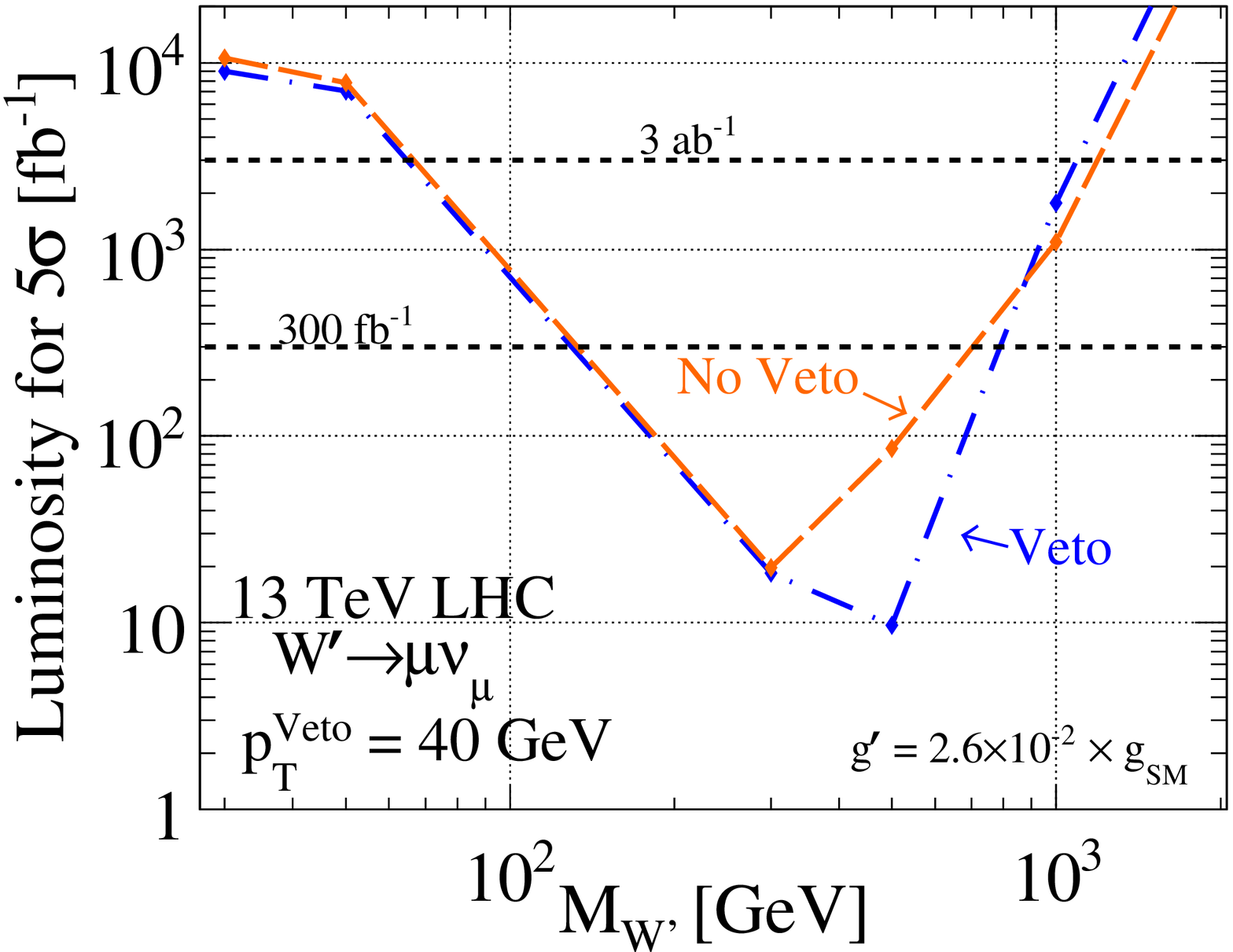}
    \label{fig:vetoPrime_Discovery}
  }
  \hspace{0.25cm}\subfigure[]{
    \includegraphics[scale=1,width=.47\textwidth]{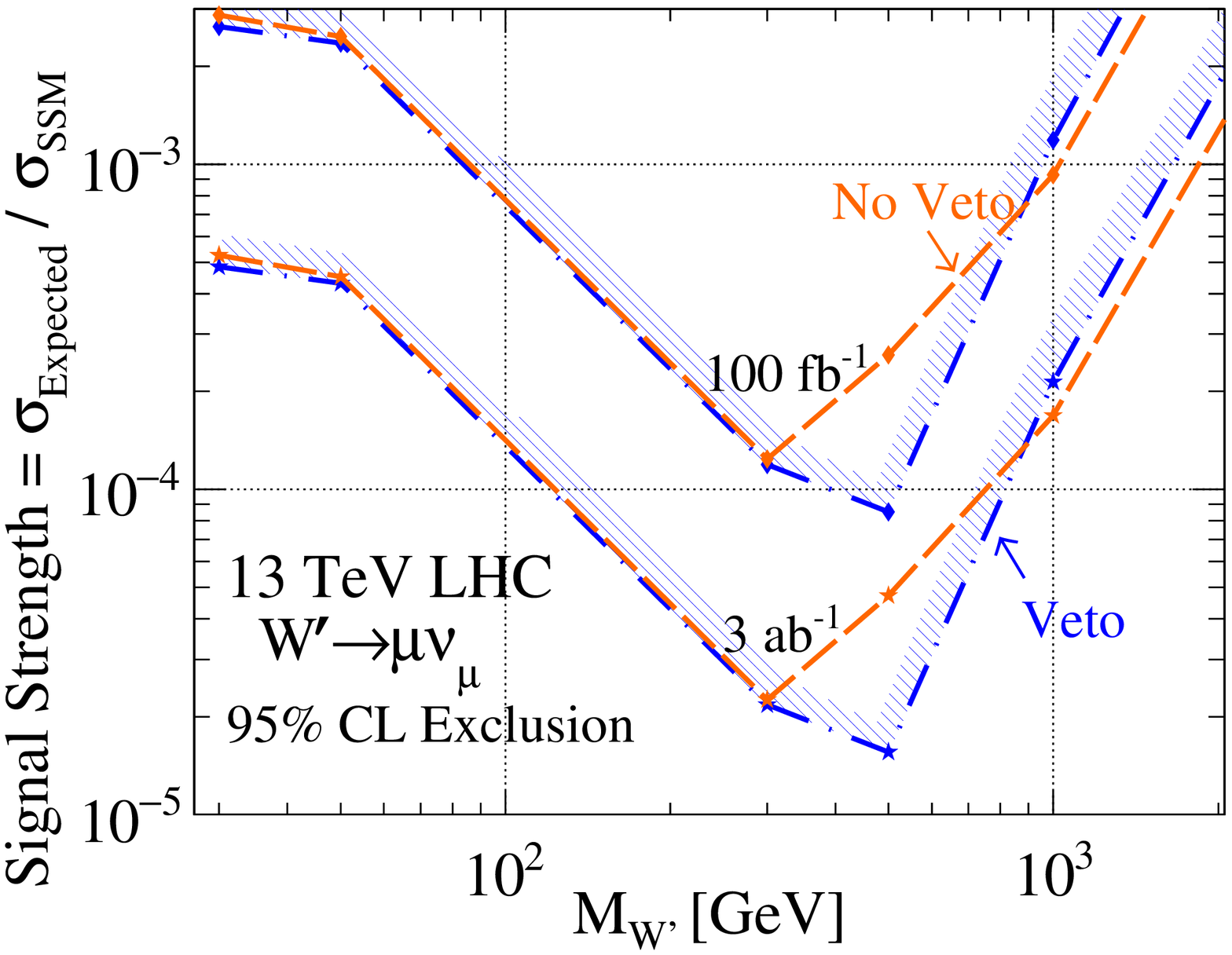}
    \label{fig:vetoPrime_Sensitivity}
  }
  \caption{Discovery potential for $W'$ boson searches via the $W'\to\mu\nu$
    channel. The results are presented in terms of the requisite luminosity for
    a $5\sigma$ statistical significance (a) of the signal over the SM
    background at the LHC (operating at a center-of-mass energy of 13~TeV) with
    (dash-dot) and without (dash) the use of a jet veto of $\pTVeto = 40\GeV$ in
    the analysis, assuming the $W'$ coupling normalization of
    Eq.~\eqref{eq:cmsCoupLimit}, as well as in terms of a 95\% confidence level
    upper limit on the signal strength (b) when assuming an integrated
    luminosity of 100 and 3000 $\invfb$.
  }
  \label{fig:discovery}
\end{figure}

To quantify the discovery potential of a positive $W'\rightarrow\mu\nu$ signal
at the LHC, we use Gaussian statistics to define the significance of a would-be
discovery as
\begin{equation}
  \sigma^{\rm Discovery} = \cfrac{n^s}{\sqrt{n^s + (1+\delta_b)n^b}}
  \qquad\text{where}\qquad n^{s,b} = \mathcal{L} \times \sigma^{s,b}\ .
\end{equation}
Here, $n^{s,b}$ represents the expected number of signal and background events given
an integrated luminosity $\mathcal{L}$ and a (fiducial) signal and background 
cross section $\sigma^{s,b}$.
Conservatively, we have introduced a $\delta_b$ parameter to account for the
potential systematic uncertainties, which we chose to be $\delta_b = 20\%$.
While for the discovery potential we require
$\sigma^{\rm Discovery}>5$, we impose $\sigma^{\rm Discovery}<2$ for
approximately evaluating the 95\% confidence level exclusion range.

\begin{table}
 \renewcommand{\arraystretch}{1.8}
 \setlength\tabcolsep{6pt}
 \centering
\begin{tabular}{ c || c | c | c | c | c | c | c  }
  $M_{W'}$ [GeV]	& 30	& 50 	& 300 	& 500 	& 1000 	 & 3000	  & 5000 \\
  \hline\hline
  $\delta(S/B)$		& 9.8\%	& 6.3\%	& 23\%	& 250\%	& -4.0\% & -5.5\% & -2.4\% \\
\end{tabular}
\caption{
Change in signal-to-noise ratio [\%] for $pp\rightarrow W'\rightarrow \mu+\MET$ searches at the 13 TeV LHC
after applying a jet veto of $\pTVeto = 40\GeV$.
}
\label{tb:snRatio}
\end{table}

In Figure~\ref{fig:vetoPrime_Discovery}, as a function of $M_{W'}$,
we show the requisite integrated luminosity for obtaining a $5\sigma$
statistical significance (or equivalently to claim discovery of a signal) 
 with (dash-dot) and without (dash) including a jet veto of $\pTVeto=40\GeV$ in the analysis. 
 As a benchmark, we assume SSM $W'$ coupling to
fermions as given in Eq.~\eqref{eq:cmsCoupLimit}, so that
$\kappa^{q,\ell}_L = g'/g_{\rm SM} = 2.6\times10^{-2}$. For light and moderate
$W'$ boson masses of $M_{W'} \in [30 , 900]\GeV$, we observe a systematic,
albeit marginally for the \comment{lighter} cases, improvement in the discovery potential.
For $M_{W'} \lesssim 100\GeV$, the signal-to-noise ratio slightly increases by
about \confirm{6-10\%} when a jet veto is employed, whereas the improvement
reaches  approximately $20\%$ and $200\%$ for heavier $W'$ boson masses of about
300~GeV or 500~GeV, respectively. This translates to
requiring $5-90\%$ less data to achieve the same $5\sigma$ statistical
sensitivity as without a jet veto for this particular mass regime. The large
variation in the utility of a jet veto is due to the relative
contribution of the top quark and DY processes in the
SM background. For small $M_{W'}$ values, non-colored-singlet backgrounds
make up only $\mathcal{O}(10\%)$ of the total background, a number that
grows dramatically for $W'$ 
\comment{scenarios above the top quark mass threshold.}
For situations in which the $W'$ boson mass is heavier than 700-800~GeV, 
the veto ceases to be useful as the SM background essentially vanishes. 
Subsequently, the veto only acts to decrease the rates leaving 
the signal-to-noise ratio unchanged, which therefore worsens the sensitivity. 
Alternatively, we show in Table~\ref{tb:snRatio} the corresponding changes 
in the signal-to-noise ratio when a jet veto is applied. 
While improvements are at the 10\% level for light $W'$ bosons, 
they drastically increase from 20\% to 250\% for moderate
$W'$ boson masses ranging from 300~GeV to 800~GeV, 
before worsening the search for the very massive $W'$ bosons.

In Figure~\ref{fig:vetoPrime_Sensitivity}, we translate this discovery
potential into a sensitivity on signal strength $\mu$ defined as
\begin{equation}
 \mu = \sigma^{\rm Expected} / \sigma^{pp\rightarrow W_{\rm SSM} \to \mu+\nu}\ ,
\end{equation}
where $\sigma^{\rm Expected}$ is the expected \comment{fiducial} signal cross section for any
$W'$ boson scenario that one may consider and
$\sigma^{pp\rightarrow W_{\rm SSM} \to \mu+\nu}$ is the \comment{analogous} SSM boson cross section
obtained when using the couplings of Eq.~\eqref{eq:cmsCoupLimit}.
We obtain similar results to (a) and
observe that jet vetoes can potentially improve the sensitivity by $5-70\%$ for
moderate $W'$ boson lying in the $300-900$~GeV mass window.

\section{Summary and Conclusion}\label{sec:summary}
The origin of tiny nonzero neutrinos masses, the particle nature of dark matter
and the weakness of gravity are longstanding issues, among others, that can
potentially be understood and studied at collider experiments via the probes for
the existence of new $W'$ and $Z'$ gauge bosons. Due to their color-singlet
nature, the QCD radiation pattern of $W'$ and $Z'$ boson production at
hadron colliders is intrinsically softer than the $W'/Z'$ mass scale and more
collinear with respect to the beam axis than the pattern associated with the leading
color non-singlet background processes. As a consequence, the sensitivity to
color-singlet new physics searches can be improved with the usage of jet
vetoes provided the QCD processes are a non-negligible fraction of the
background.

As a proof of principle, we have studied at the 13 TeV LHC, the muonic signature of a generic $W'$ signal,
\begin{equation}
p p \rightarrow W' \rightarrow \mu \nu \ ,
\end{equation}
 focusing on the increased discovery potential gained by
employing jet vetoes. We have systematically considered both signal and
background processes at NLO+PS accuracy, and included, for color-singlet signal
and background channels, the resummation of jet veto logarithms up to the NNLL
accuracy with matching to NLO fixed-order results.
This has necessitated the development of a new {\sc FeynRules} model
in which we have implemented in a generic fashion 
new $W'$ and $Z'$ gauge bosons with model-independent chiral couplings.
See Section~\ref{sec:theory} for more details.
Associated model files are public available from the {\sc FeynRules}
model database~\cite{WprimeZprimeAtNLO}.

We have investigated the impact of several classes of uncertainties that
are attached to jet veto resummation calculations. We have probed the
dependence on the choice of jet \comment{definition, 
which suggests larger jet radii $(R\sim1)$ lead to smaller uncertainties than smaller radii.}
See Section~\ref{sec:phenoModeling} for additional details.

In Section~\ref{sec:metJetModeling}, 
we studied the dependence of our collider analysis on missing transverse energy definitions
as well as the flavor-dependence of the jet vetoes
We have described how exclusive missing energy definitions, such as the one used in 13~TeV ATLAS analyses, 
broaden all missing-energy-based observables and subsequently leads to a decrease in experimental sensitivity. 
This choice of $\MET$ additionally leads to the rise of a new class of non-global logarithms that are responsible 
for a \comment{potentially} large theoretical uncertainties that have not been previously taken into account.
On different lines, we have found that with respect to a $b$-jet veto of $\pTVeto=30\GeV$, flavor-agnostic jet
vetoes of $\pTVeto = 30-40\GeV$  can further reduce single top and
top-antitop quark production by a factor of \confirm{$2-50$}  at a mild cost of the signal rate.

We have applied our finding to the specific case of a $\Wssm$ boson, and
observed that for a new physics coupling strength taken as large as allowed by
the current constraints, $\kappa^{q,\ell}_L = g'/g_{\rm SM} = 2.6\times10^{-2}$.
The usage of jet vetoes can increase the signal-to-noise ratios by
roughly 10\% for very light bosons masses of $30-50$~GeV and
25\%-250\% for moderately heavy bosons of $300-800$~GeV.
Beyond this, vetoes lose there usefulness as they decrease the signal rates by a large
amount, leaving the almost vanishing SM backgrounds almost unaffected.
Conversely, $\Wssm$-bosons could be discovered by using 2-10 times for moderate $M_{W'}$,
the moderate mass range, in contrast to any other mass scale where the change is milder.

\section*{Acknowledgements}

Thomas Becher, Lydia Brenner, Fabrizio Caola, Mihoko Nojiri, Cedric Weiland are thanked for valuable discussions. 
The work of RR was funded in part by the UK Science and Technology Facilities Council (STFC), 
the European Union's Horizon 2020 research and innovation programme under the Marie Sklodowska-Curie grant agreement No 690575,
and Grant-in-Aid for Scientific Research from the Ministry of Education, Science, Sports, and Culture (MEXT), 
Japan (Nos.~16H06492 and 16H03991 for M.~M.~Nojiri).
The authors appreciate the hospitality of IHEP where this work was initiated.
RR acknowledges the generous hospitality KEK and the European Union.

\bibliographystyle{JHEP}


\end{document}